\documentclass[sigconf]{acmart}

\copyrightyear{2022}
\acmYear{2022}
\setcopyright{acmcopyright}\acmConference[CCS '22]{Proceedings of the 2022 ACM
SIGSAC Conference on Computer and Communications Security}{November 7--11,
2022}{Los Angeles, CA, USA}
\acmBooktitle{Proceedings of the 2022 ACM SIGSAC Conference on Computer and
Communications Security (CCS '22), November 7--11, 2022, Los Angeles, CA, USA}
\acmPrice{15.00}
\acmDOI{10.1145/3548606.3560697}
\acmISBN{978-1-4503-9450-5/22/11}

\usepackage{algorithm}
\makeatletter
\renewcommand{\ALG@name}{Protocol}
\makeatother
\usepackage{algorithmic}
\usepackage{graphicx}
\graphicspath{./pictures/}
\usepackage[algo2e]{algorithm2e}
\usepackage{amsmath, amsthm}
\usepackage{multirow}
\usepackage{dcolumn}
\newlength{\thinline}
\newlength{\thickline}
\usepackage{listings}
\usepackage{url}
\newtheorem{theorem}{Theorem}

\usepackage{subfigure}
\usepackage{balance}
\usepackage{enumitem}
\usepackage{hyperref}

\newcounter{protocol}

\newcommand{\pmpl}{\texttt{pMPL}\xspace}

\begin{document}
\begin{sloppypar}

\title{\texttt{pMPL}: A Robust Multi-Party Learning Framework with a Privileged Party}

\author{
Lushan Song
}
\affiliation{Fudan University}
\email{19110240022@fudan.edu.cn}

\author{
Jiaxuan Wang
}
\affiliation{Fudan University}
\email{20212010090@fudan.edu.cn}

\author{
 Zhexuan Wang
}
\affiliation{Fudan University}
\email{21210240331@m.fudan.edu.cn}

\author{
Xinyu Tu
}
\affiliation{Fudan University}
\email{21210240326@m.fudan.edu.cn}

\author{
Guopeng Lin
}
\affiliation{Fudan University}
\email{17302010022@fudan.edu.cn}

\author{
Wenqiang Ruan
}
\affiliation{Fudan University}
\email{20110240031@fudan.edu.cn}

\author{
Haoqi Wu
}
\affiliation{Fudan University}
\email{19212010008@fudan.edu.cn}

\author{
Weili Han
}
\affiliation{Fudan University}
\email{wlhan@fudan.edu.cn}

\renewcommand{\shortauthors}{Lushan Song et al.}

\begin{abstract}
In order to perform machine learning among multiple parties while protecting the privacy of raw data, privacy-preserving machine learning based on secure multi-party computation (MPL for short) has been a hot spot in recent. The configuration of MPL usually follows the peer-to-peer architecture, where each party has the same chance to reveal the output result. However, typical business scenarios often follow a hierarchical architecture where a powerful, usually \textit{privileged party}, leads the tasks of machine learning. Only the \textit{privileged party} can reveal the final model even if other \textit{assistant parties} collude with each other. It is even required to avoid the abort of machine learning to ensure the scheduled deadlines and/or save used computing resources when part of \textit{assistant parties} drop out.

Motivated by the above scenarios, we propose \pmpl, a robust MPL framework with a \textit{privileged party}. \pmpl supports three-party (a typical number of parties in MPL frameworks) training in the semi-honest setting. By setting alternate shares for the \textit{privileged party}, \pmpl is robust to tolerate one of the rest two parties dropping out during the training. 
With the above settings, we design a series of efficient protocols based on vector space secret sharing for \pmpl to bridge the gap between vector space secret sharing and machine learning.
Finally, the experimental results show that the performance of \pmpl is promising when we compare it with the state-of-the-art MPL frameworks. Especially, in the LAN setting, \pmpl is around $16\times$ and $5\times$ faster than \texttt{TF-encrypted} (with \texttt{ABY3} as the back-end framework) for the linear regression, and logistic regression, respectively.
Besides, the accuracy of trained models of linear regression, logistic regression, and BP neural networks can reach around 97\%, 99\%, and 96\% on MNIST dataset respectively.

\end{abstract}



\maketitle

\section{Introduction}
\label{sec.introduction}

Privacy-preserving machine learning based on secure multi-party computation (MPC for short), referred to as secure multi-party learning (MPL for short)~\cite{songsok}, allows multiple parties to jointly perform machine learning over their private data while protecting the privacy of the raw data. 
MPL breaks the barriers that different organizations or companies cannot directly share their private raw data  mainly due to released privacy protection regulations and laws~\cite{DBLP:journals/ieeesp/RuanXJWSH21} (e.g. GDPR~\cite{voigt2017eu}). Therefore, MPL can be applied to several practical fields involving private data, such as risk control in the financial field~\cite{DBLP:conf/kdd/0001ZWWFTWLWH21} and medical diagnosis~\cite{esteva2017dermatologist,fakoor2013using}.

Researchers have proposed a doze of MPL frameworks~\cite{byali2020flash,mohassel2018aby3,mohassel2017secureml,DBLP:conf/ndss/ChaudhariRS20,koti2021swift,dalskov2021fantastic,wagh2019securenn}, which support $\ge$2 computation parties during the learning. 
The involved parties usually follow the peer-to-peer architecture according to the protocols that they rely on. That is, each of them has the same chance to handle the results, including intermediate results and the final model after training. In \texttt{ABY3}~\cite{mohassel2018aby3}, for example, any two parties can cooperate with each other to obtain the final model after training. However, it is also necessary to provide a hierarchical architecture, where a party has its privileged position to handle the process and results of learning due to its motivation and possible payments (including computing resources, and money), in practical scenarios.

\vspace{-2mm}
\subsection{Practical Scenarios}
\label{sec.scenario}

\begin{figure}[ht]
\centering
\vspace{-4mm}
\includegraphics[scale=0.65]{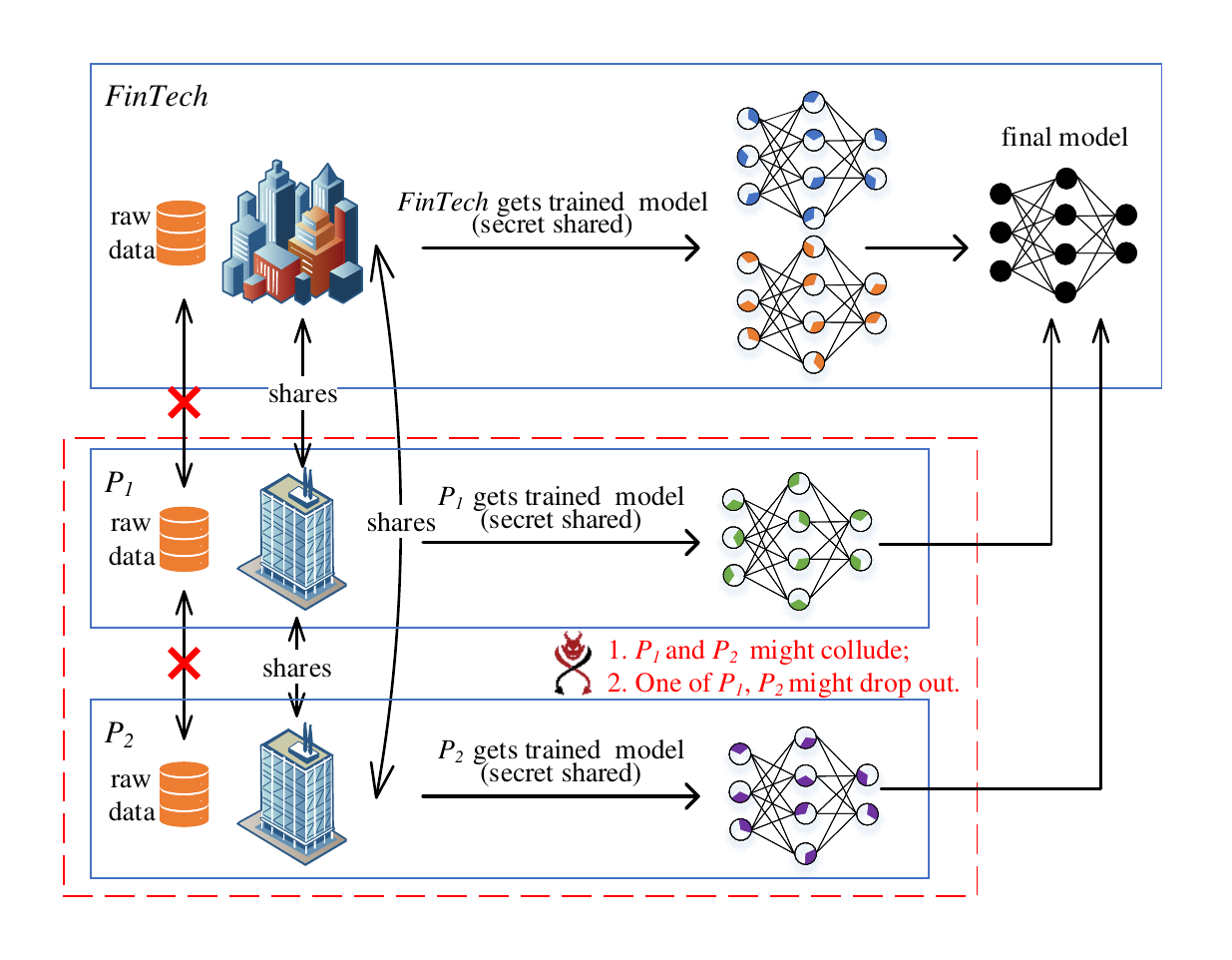}
\vspace{-6mm}
\caption{Practical scenarios}
\label{fig.scenario}
\vspace{-6mm}
\end{figure}

As is shown in Figure~\ref{fig.scenario}, three parties, i.e. \textit{FinTech}, $P_1$ and $P_2$,  are involved in a scenario of the financial risk control: \textit{FinTech} is a professional company (usually with a big volume of authorized data and capital) in the financial field. While $P_1$ and $P_2$ are two Internet service providers, which usually have lots of valued data (with authorization from their users).  \textit{FinTech} wants to cooperate with  $P_1$ and $P_2$  to train an accurate model for the financial risk control, under the payments for the data, which are used in the training process, from $P_1$ and $P_2$. However, \textit{FinTech}, $P_1$ and $P_2$ cannot exchange the raw data with each other due to the released privacy protection regulations and laws (e.g. GDPR~\cite{voigt2017eu}). Besides, one party could suffer system or network failures, or intentionally quit the training process of machine learning for business purposes, e.g. requiring more payments. Thus, the proposed framework should tolerate the dropping out of a party ($P_1$ or $P_2$). For the former case, although parties could restart the training process to deal with the dropping, it should be more practical that the training process is continued to the end, because it can ensure the scheduled deadlines and/or save used computing resources. For the latter case, the proposed framework must support continuing the secure joint  training only with the rest parties.
 
In the above scenario, \textit{FinTech} requires a privileged position under the payments: (1) \textit{FinTech} is the only party to reveal the final model, even when $P_1$ and $P_2$ collude with each other; (2) After being launched, the training process can be continued to the end, even when $P_1$ or $P_2$ drops out due to objective or subjective reasons. Note that \textit{FinTech} can leverage the robustness to choose one party to reveal the final model, thus keeping its privileged position until the end of training.
With the privileged position, \textit{FinTech} will be much more motivated and responsible to deploy MPL frameworks among parties. Thus, the hierarchical architecture is necessary for the development of the studies of MPL frameworks.

As is shown in Figure~\ref{fig.scenario}, three parties, i.e. \textit{FinTech}, $P_1$ and $P_2$, hold shares rather than raw data to train models with the support of a series of MPC protocols. After the training, $P_1$ and $P_2$ send their shares of the trained model to \textit{FinTech} to ensure that \textit{FinTech} is the sole one to reveal the final model. Note that $P_1$ and $P_2$ cannot reveal the final model even by colluding with each other. Furthermore, for the second requirement, after three parties hold shares, the training process can be continued with shares of \textit{FinTech}+ $P_1$ or \textit{FinTech}+ $P_2$ if  $P_2$ or  $P_1$ drops out.

\vspace{-2mm}
\subsection{Related Work}

Privacy-preserving machine learning, especially based on MPC technologies, has become a hot spot in recent years. Researchers have proposed a doze of MPL frameworks~\cite{byali2020flash,mohassel2018aby3,mohassel2017secureml,DBLP:conf/ndss/ChaudhariRS20,koti2021swift,dalskov2021fantastic,wagh2019securenn}.

Several MPL frameworks were designed based on additive secret sharing~\cite{DBLP:conf/esorics/BogdanovLW08}.
For instance, Mohassel and Zhang~\cite{mohassel2017secureml} proposed a two-party MPL framework, referred to as \texttt{SecureML}, which supported the training of various machine learning models, including linear regression, logistic regression, and neural networks.
Wagh et al.~\cite{wagh2019securenn} designed a three-party MPL framework \texttt{SecureNN} based on additive secret sharing. They eliminated expensive cryptographic operations for the training and inference of neural networks.
In the above MPL frameworks, the training would be aborted if one party dropped out.

In addition, a majority of MPL frameworks were designed based on replicated secret sharing~\cite{DBLP:conf/ccs/ArakiFLNO16}.
Mohassel and Rindal~\cite{mohassel2018aby3} proposed \texttt{ABY3}, a three-party MPL framework. It supported efficiently switching back and forth among arithmetic sharing~\cite{DBLP:conf/esorics/BogdanovLW08}, binary sharing~\cite{DBLP:conf/stoc/GoldreichMW87}, and Yao sharing~\cite{DBLP:conf/ccs/MohasselRZ15}.
\texttt{Trident}~\cite{DBLP:conf/ndss/ChaudhariRS20} extended \texttt{ABY3} to four-party scenarios, and outperformed it in terms of the communication complexity.
In both \texttt{ABY3} and \texttt{Trident}, any two parties can corporate to reveal the secret value (e.g. the final model after training). Therefore, \texttt{ABY3} and \texttt{Trident} can ensure the robustness that tolerated one of the parties dropping out in the semi-honest security model.
Furthermore, several MPL frameworks~\cite{byali2020flash,koti2021swift,dalskov2021fantastic} were designed to tolerate the dropping out of one malicious party during training. That is, even though there existed a malicious party, these MPL frameworks can still continue training, and produce correct outputs.
\texttt{FLASH}~\cite{byali2020flash} and \texttt{SWIFT}~\cite{koti2021swift} assumed that there existed one malicious party and three honest parties. They ensured robustness by finding an honest party among four parties, and delegating the training to it.
\texttt{Fantastic Four}~\cite{dalskov2021fantastic} assumed there existed one malicious party and three semi-honest parties. It ensured the robustness by excluding the malicious party, and the rest parties can continue training securely.
Note that the approaches of \texttt{FLASH} and \texttt{SWIFT} would leak the sensitive information of other parties to the honest party, while \texttt{Fantastic Four} would not leak the sensitive information during training. However, any two parties of \texttt{Fantastic Four} (including  \texttt{FLASH} and \texttt{SWIFT}) can corporate to reveal the final results. In summary, \texttt{Fantastic Four} cannot set a \textit{privileged party} because it followed a peer-to-peer architecture.

The existing MPL frameworks~\cite{byali2020flash,mohassel2018aby3,mohassel2017secureml,DBLP:conf/ndss/ChaudhariRS20,koti2021swift,dalskov2021fantastic,wagh2019securenn} cannot meet both two requirements mentioned above, although these two ones are important in practical scenarios.
For MPL frameworks~\cite{mohassel2017secureml,wagh2019securenn} based on additive secret sharing, they can only meet the first requirement, while cannot meet the second one because when one of the \textit{assistant parties} drops out during training, the machine learning tasks will be aborted.
At the same time, several MPL frameworks~\cite{byali2020flash,mohassel2018aby3,DBLP:conf/ndss/ChaudhariRS20,koti2021swift,dalskov2021fantastic} based on replicated secret sharing have such robustness in the second requirement, while cannot meet the first one, because the final results can be revealed by the cooperation of any $t$ ($\le$n) parties. That is, these frameworks follow the peer-to-peer architecture.

In addition to MPL, federated learning~\cite{xu2019verifynet,konevcny2016federated1,konevcny2016federated2} and trusted execution environments~\cite{ohrimenko2016oblivious} are two other paradigms of privacy-preserving machine learning.
In federated learning, each client trains a model with its owned data locally, and uploads the model updates rather than the raw data to a centralized server. Although federated learning has a relatively higher efficiency than that of MPL frameworks, the model updates might contain sensitive information, which might be leaked~\cite{melis2019exploiting,DBLP:series/lncs/Zhu020} to the server and other involved clients.
In addition, in federated learning, Shamir's secret sharing~\cite{DBLP:journals/cacm/Shamir79} can be used to ensure the robustness that tolerates part of clients dropping out during the training~\cite{DBLP:conf/ccs/BonawitzIKMMPRS17}. The differences between federated learning and our proposed framework will be discussed in Section~\ref{sec.flmpl}.
For trusted execution environments, they train models over a centralized data source from distributed locations based on extra trusted hardware. The security model has one or several third trusted parties, thus significantly differs from those of MPL frameworks.
The privacy is preserved by the trustworthiness of the data process environment, where parties only obtain the final results without knowing the details of raw data.

\vspace{-2mm}
\subsection{Our Contributions}
\label{sec.contribution}

In this paper, we are motivated to leverage the vector space secret sharing~\cite{brickell1989some}, which is typically applied in the cryptographic access control field, to meet the above requirements. Based on vector space secret sharing, we propose a robust MPL framework with a \textit{privileged party}, referred to as \pmpl\footnote{We open our implementation codes at GitHub (https://github.com/FudanMPL/pMPL).}. Given an access structure on a set of parties, the vector space secret sharing guarantees that only the parties in the preset authorized sets can reveal the secret value shared between/among parties. Thus, we set each authorized set to include the \textit{privileged party} mentioned above, and once training is completed, only \textit{assistant parties} send their shares to the \textit{privileged party}, while the \textit{privileged party} does not send its shares to them. Therefore,  \pmpl can meet the first requirement.
To ensure the robustness mentioned in the second requirement, we let the \textit{privileged party} hold redundant shares to continue the machine learning when one \textit{assistant party} drops out.
Despite the above configuration, how to apply the vector space secret sharing to machine learning, including the technical issues of framework design, efficient protocols, and performance optimizations, is still highly challenging.

We highlight the main contributions in our proposed \pmpl as follows:

\begin{itemize}[leftmargin=*]
\item \textbf{A robust three-party learning framework with a \textit{privileged party}.}  We propose \pmpl, a three-party learning framework based on vector space secret sharing with a privileged party.
\pmpl guarantees that only the \textit{privileged party} can obtain the final model even when two \textit{assistant parties} collude with each other.  Meanwhile, \pmpl is robust, i.e. it can tolerate either of the \textit{assistant parties} dropping out during training.  
To the best of our knowledge, \pmpl is the first framework of privacy-preserving machine learning based on vector space secret sharing.

\item \textbf{Vector space secret sharing based protocols for \pmpl.} 
Based on the vector space secret sharing, we propose several fundamental efficient protocols required by machine learning in \pmpl, including secure addition, secure multiplication, secure conversion between vector space secret sharing and additive secret sharing, secure truncation. Furthermore, to efficiently execute secure multiplication, we design the vector multiplication triplet generation protocol in the offline phase.  
\end{itemize}

\noindent \textbf{Implementation}:
Our framework \pmpl can be used to train various typical machine learning models, including linear regression, logistic regression, and BP neural networks. We evaluate \pmpl on MNIST dataset. 
The experimental results show that the performance of \pmpl is promising compared with the state-of-the-art MPL frameworks, including \texttt{SecureML} and \texttt{TF-Encrypted}~\cite{DBLP:journals/corr/abs-1810-08130} (with \texttt{ABY3}~\cite{mohassel2018aby3} as the back-end framework). Especially, in the LAN setting, \pmpl is around $16\times$ and $5\times$ faster than \texttt{TF-encrypted} for the linear regression and logistic regression, respectively.
 In the WAN setting, although \pmpl is slower than both \texttt{SecureML} and \texttt{TF-encrypted}, the performance is still promising. In \pmpl, to provide more security guarantees (i.e., defending the collusion of two \textit{assistant parties}) and ensure robustness, \pmpl requires more communication overhead.  Besides, the accuracy of trained models of linear regression, logistic regression, and BP neural networks can reach around 97\%, 99\%, and 96\% on MNIST dataset, respectively. Note that the accuracy evaluation experiments of linear regression and logistic regression execute the binary classification task, while the evaluation experiments of BP neural networks execute the ten-class classification task.

\smallskip

\vspace{-2mm}
\section{Preliminaries}
\label{sec.preliminaries}
In this section, we introduce the background knowledge of MPC technologies and three classical machine learning models supported by \pmpl.

\vspace{-2mm}
\subsection{Secure Multi-Party Computation}
MPC provides rigorous security guarantees and enables multiple parties, which could be mutually distrusted, to cooperatively compute a function while keeping the privacy of the input data.
It was firstly introduced by Andrew C. Yao in 1982, and originated from the millionaires' problem~\cite{yao1982protocols}. 
After that, MPC is extended into a general definition for securely computing any function with polynomial time complexity~\cite{yao1986generate}.
Various MPC protocols, such as homomorphic encryption-based protocols~\cite{DBLP:conf/acns/GiacomelliJJPY18}, garbled circuit-based protocols~\cite{rouhani2018deepsecure}, and secret sharing-based protocols~\cite{DBLP:conf/esorics/BogdanovLW08} have their specific characteristics, and are suitable for different scenarios.

Secret sharing, which typically works over integer rings or prime fields, has proven its feasibility and efficiency in privacy-preserving machine learning frameworks~\cite{wagh2019securenn,koti2021swift,byali2020flash}. 
These frameworks are essentially built on additive secret sharing or replicated secret sharing~\cite{DBLP:conf/ccs/ArakiFLNO16}, where the secret value for sharing is randomly split into several shares, the sum of these shares is equal to the secret value.
Shamir's secret sharing~\cite{DBLP:journals/cacm/Shamir79} is another important branch of secret sharing.
In Shamir's secret sharing, the shares are constructed according to a randomized polynomial, and the secret value can be reconstructed by solving this polynomial with Lagrange interpolation.

According to the brief analysis of the two requirements of \pmpl in Section~\ref{sec.introduction}, neither two types 
of secret sharing mentioned above can meet the both requirements, i.e. supporting a \textit{privileged party} and tolerating that part of \textit{assistant parties} dropping out. 
Therefore, in our proposed \pmpl, we employ the vector space secret sharing~\cite{brickell1989some}, another
type of secret sharing, to meet the both two requirements.

\vspace{-2mm}
\subsection{Vector Space Secret Sharing} 
Vector space secret sharing~\cite{brickell1989some} can set which parties can cooperate to reveal the secret value, and which parties cannot reveal the secret value even if they collude with each other.

Let $\mathcal{P} = \{P_0, P_1, \dots, P_n \}$ be a set of parties ($P_i$ refers to the $i$-\textit{th} party), and $\varGamma = \{B_0, B_1, \dots, B_k \}$ be a set of subsets of $\mathcal{P}$, i.e. $\varGamma \subseteq 2^\mathcal{P}$. $\varGamma$ is defined as an access structure on $\mathcal{P}$. Meanwhile, its element $B_j\in\varGamma$ is defined as a authorized set in which parties can cooperate with each other to reveal the secret value. In contrast, the set of parties that is not in the access structure $\varGamma$ cannot reveal the secret value. Then, with a large prime number $p$ and an integer $d$ where $d \ge 2$, we notify $(\mathbb{Z}_p)^d$ as the vector space over $\mathbb{Z}_p$.
Suppose there is a function $\varPhi: \mathcal{P} \to (\mathbb{Z}_p)^d$ that satisfies the following property:

\begin{equation}
\begin{aligned}
    \label{property}
    & (1,0, \dots, 0)~ \text{\textit{can be written as a linear combination of }} \\  
    &  \text{\textit{ elements in the set}}~\{ \varPhi(P_i) ~ | ~ P_i \in B_j \}
    \Leftrightarrow B_j \in \varGamma 
\end{aligned}
\end{equation}

That is, for any authorized set $B_j$, $(1,0, \dots, 0)$ can be represented linearly by all the \textit{public vectors} in the set $\{ \varPhi(P_i) ~ | ~ P_i \in B_j \} $.  Therefore, there are $m$ public constants $c_0,...,c_{m-1}$ (we name them as reconstruction coefficients in this paper), where $m$ refers to the number of parties in $B_j$, such that: 
\begin{equation}
    (1,0, \dots, 0) = \sum_{P_i \in B_j} c_i \cdot {\varPhi(P_i)}
\end{equation}

We denote the matrix constructed by the \textit{public vectors} as $\varPhi (\mathcal{P})$, and name it the \textit{public matrix}.
Suppose that the \textit{public matrix} $\varPhi(\mathcal{P})$ has been determined by all the parties. To secret share a value $x$, the party who holds this value samples $d-1$ random values $s_1, s_2, \dots, s_{d-1} \in \mathbb{Z}_p$. Then it constructs the vector $\vec{s} = (x, s_1, s_2, \dots, s_{d-1})^T$. After that, this party computes the share $x_i = \varPhi(P_i) \times \vec{s}$ corresponding to $P_i$, where $0 \leq i \leq n$.

According to the above share generation mechanism, we can observe that $(1,0, \dots, 0) \times \vec{s} = x$. Hence:
\begin{equation}
\begin{aligned}
\label{eq.rec}
    x &= \Big(\sum_{P_i \in B_j} c_i \cdot \varPhi(P_i) \Big) \times \vec{s}  = \sum_{P_i \in B_j} c_i \cdot x_i, B_j \in \varGamma
\end{aligned}
\end{equation}

Therefore, parties can reveal the secret value $x$ by computing Equation (\ref{eq.rec}).

\vspace{-2mm}
\subsection{Machine Learning Models}
\label{sec.ml}
We introduce three typical machine learning models supported by \pmpl as follows:

\noindent\textbf{Linear Regression:}  
With a matrix of training samples $\mathbf{X}$ and the corresponding vector of label values $\mathbf{Y}$, linear regression learns a function $G$, such that $G(\mathbf{X}) = \mathbf{X} \times \vec{w} \approx \mathbf{Y}$, where $\vec{w}$ is a vector of coefficient parameters.
The goal of linear regression is to find the coefficient vector $\vec{w}$ that minimizes the difference between the output of function $G$ and label values. 
The forward propagation stage in linear aggression is to compute $\mathbf{X} \times \vec{w}$. Then, in the backward propagation stage, the coefficient parameters $\vec{w}$ can be updated as :
    \begin{equation}
    \label{eq.linear_w}
    \vec{w} :=\vec{w}-\alpha \mathbf{X}^{T}(\mathbf{X} \times \vec{w}-\mathbf{Y})
    \end{equation}
where $\alpha$ is the learning rate.

\noindent\textbf{Logistic Regression:} 
In binary classification problems,  
logistic regression introduces the logistic function $f(u) = \frac{1}{1+e^{-u}}$ to bound the output of the prediction between 0 and 1. Thus the relationship of logistic regression is expressed as $G(\mathbf{X}) = f(\mathbf{X} \times \vec{w})$.
The forward propagation stage in logistic regression is to compute $f(\mathbf{X} \times \vec{w})$. Then, in the backward propagation stage, the coefficient parameters $\vec{w}$ can be updated as:
    
    \begin{equation}
    \label{eq.logistic_w}
    \vec{w} :=\vec{w}-\alpha \mathbf{X}^{T}(f(\mathbf{X} \times \vec{w})-\mathbf{Y})
    \end{equation}

\noindent\textbf{BP Neural Networks:} 
Back propagation (BP for short) neural networks can learn non-linear relationships among high dimensional data. A typical BP neural network consists of one input layer, one output layer, and multiple hidden layers. 
Each layer contains multiple nodes, which are called neurons. Except for the neurons in the input layer, each neuron in other layers comprises a linear function, followed by a non-linear activation function $f(u)$ (e.g. \texttt{ReLu}). 
In addition, neurons in the input layer take training samples as the input, while other neurons receive their inputs from the previous layer, and process them to produce the computing results that serve as the input to the next layer.

We denote the input matrix as $\mathbf{X}_0$, the coefficient matrix of the $(i-1)$-\textit{th} layer to the $i$-\textit{th} layer as $\mathbf{W}_i$ and the output matrix as $\mathbf{Y}_m$. 
In the forward propagation stage in BP neural networks, the output of the $i$-th layer is computed as $\mathbf{A}_i = f(U_i)$, where $\mathbf{U}_i = \mathbf{A_{i-1}} \times \mathbf{W}_i$, and $f(\cdot)$ is the activation function of the $i$-th layer. In addition, $\mathbf{A}_0$ is initialized as $\mathbf{X}_0$, and the output matrix is $\mathbf{A}_m$. In the backward propagation stage, the error matrix for the output layer is computed as $\mathbf{E}_m=(\mathbf{A}_m-\mathbf{Y}_m)$, and the error matrices of other layers are computed as  $\mathbf{E}_i = (\mathbf{E}_{i+1} \times \mathbf{W}_i^T) \odot \partial f(\mathbf{U}_i)$. Here $\odot$ denotes the element-wise product, and $\partial f(\cdot)$ denotes the derivative of activation function $f(\cdot)$. After the backward propagation phase, we update the coefficient matrix as $\mathbf{W}_i := \mathbf{W}_i - \alpha \mathbf{A}_{i-1}^T \times \mathbf{E}_i$.

\section{Overview of \pmpl}
\label{sec.overview}

In this section, we firstly describe the architecture of \pmpl, and introduce the data representation of \pmpl. After that, we present the security model considered in this paper. Finally, we introduce the design of robust training of \pmpl. 
For the clarity purpose, we show the notations used in this paper in Table~\ref{table.notation}.

\vspace{-2mm}
\begin{table}[h]
\caption{Notations used in this paper.}
\label{table.notation}
\renewcommand\arraystretch{1}
\vspace{-2mm}
\scalebox{0.8}{
\begin{tabular}{ll}
\toprule
Symbol                                                      & Description                                                                                                          \\ \midrule
$\mathcal{P}$                                               & The set of parties                                                                                                   \\ 
$\varGamma$                                                 & The access structure                                                                                                 \\ 
$B_j$                                                       & The authorized set                                                                                         \\ 
$[ \cdot ]$                                                 & The shares of additive secret sharing                                                                                \\ 
$[ \cdot ] ^2$                                              & The shares of  boolean sharing                                                                                       \\ 
$\langle \cdot \rangle$                                     & The shares of vector space secret sharing                                                                            \\ 
$\varPhi (\mathcal{P})$                                                   & The \textit{public matrix} for vector space secret sharing                                                           \\ 
$c_0, c_1, \dots, c''_3$                                    & The reconstruction coefficients                                                                                       \\ 
$a_0, a_1$                                                  & The coefficients of the alternate vector                                                                             \\ 
$\ell$                                                      & The number of bits to represent a fixed-point number                                                                 \\ 
$\ell_f$                                                    & \begin{tabular}[c]{@{}l@{}}The number of bits to represent the fractional \\ part of a fixed-point number\end{tabular} \\ 
${\langle u \rangle, \langle v \rangle, \langle h \rangle}$ & The vector multiplication triplet                                                                                    \\ 
$B$                                                         & The batch size                                                                                                       \\ 
$D$                                                         & The dimension of the feature                                                                                         \\ 
$E$                                                         & The number of the epoch                                                                                         \\ \bottomrule
\end{tabular}}
\vspace{-2mm}
\end{table}

\vspace{-2mm}
\subsection{Architecture and Data Representation}
\label{sec.frameworkanddp}
\subsubsection{Architecture}~\label{sec.framework} 
\begin{figure}[ht]
\centering
\vspace{-6mm}
\includegraphics[scale=0.60]{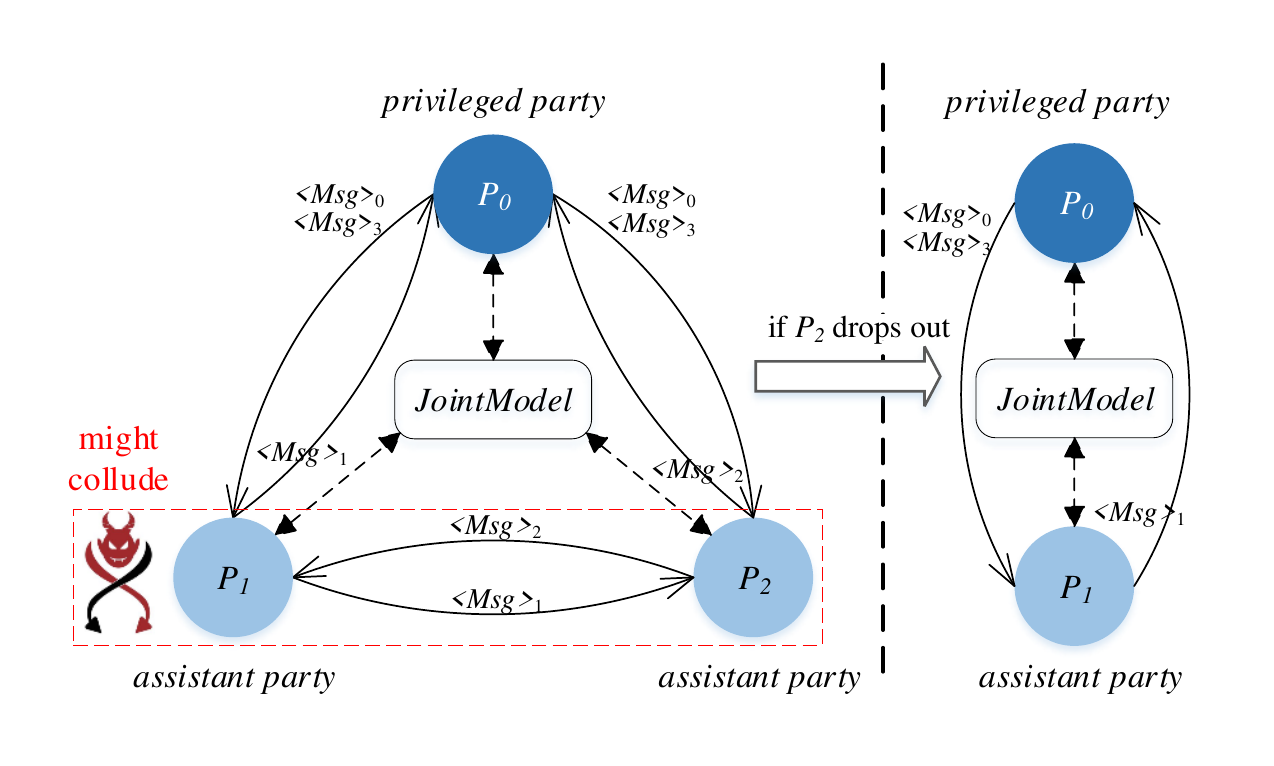}
\vspace{-6mm}
\caption{Overview of \pmpl}
\label{fig.architecture}
\vspace{-6mm}
\end{figure}

As is shown in Figure~\ref{fig.architecture}, we consider a set of three parties $\mathcal{P}=\{P_0, P_1, P_2\}$, who want to train various machine learning models over their private raw data jointly. Without loss of generality, we define $P_0$ as the \textit{privileged party} and $P_1, P_2$ as \textit{assistant parties}. These parties are connected by secure pairwise communication channels in a synchronous network.
Before training, these parties secret share (using the $\langle \cdot \rangle$\textit{-sharing} semantics introduced in Section \ref{sec.ss}) their private raw data with each other. 
During training, all the parties communicate the shared form $\langle \mathit{Msg} \rangle$ of intermediate messages with each other. 
In \pmpl, the \textit{privileged party} $P_0$ holds $\langle \mathit{Msg} \rangle_0$ and $\langle \mathit{Msg} \rangle_3$, and \textit{assistant parties} $P_1$ and $P_2$ hold $\langle \mathit{Msg} \rangle_1$ and $\langle \mathit{Msg} \rangle_2$ respectively.
During the training process, none of the parties can get others' raw data or infer any private information from the intermediate results and the final model.

Besides, the final model is supposed to be obtained only by \textit{privileged party} $P_0$, even when $P_1$ and $P_2$ collude with each other. 
Furthermore, \pmpl tolerates one \textit{assistant party} ($P_1$ or $P_2$) dropping out of training. As a result, the access structure $\varGamma$ in \pmpl is $\{ \{ P_0, P_1, P_2 \}, \{ P_0, P_1 \}, \{ P_0, P_2 \} \}$.

\subsubsection{Data representation} 

In machine learning, to train accurate models, most of the intermediate values are represented as floating-point numbers. However, since the precision of floating-point numbers is not fixed, every calculation requires additional operations for alignment. Therefore, floating-point calculations would lead to more computation and communication overhead.

In order to balance the accuracy and efficiency of the floating-point calculations in \pmpl, we handle floating-point values with a fixed-point representation.
More specifically, we denote a fixed-point decimal as an $\ell$-bit integer, which is identical to the previous MPL frameworks (e.g. \texttt{SecureML}~\cite{mohassel2017secureml}).
Among these $\ell$ bits, the most significant bit (MSB) represents the sign and the $\ell_f$ least significant bits are allocated to represent the fractional part. 
An $\ell$-bit integer can be treated as an element of a ring $\mathbb{Z}_{2^\ell}$.
Note that to ensure that corresponding reconstruction coefficients can be computed for any \textit{public matrix}, vector space secret sharing usually performs on a prime field. 
However, it is more efficient to work on a ring~\cite{DBLP:conf/sp/Damgard0FKSV19}. Therefore, we perform our computations on a ring $\mathbb{Z}_{2^\ell}$ by restricting the \textit{public matrix} (see Section~\ref{sec:share_reveal} for more detail).

\vspace{-2mm}
\subsection{Security Model}
In this paper, we employ the semi-honest (also known as honest-but-curious or passive) security model in \pmpl.
A semi-honest adversary attempts to infer as much information as possible from the messages they received during training. However, they follow the protocol specification.
Furthermore, we have an asymmetric security assumption that \textit{assistant parties} $P_1$ and $P_2$ might collude, and the \textit{privileged party} $P_0$ would not collude with any \textit{assistant party}. This setting is different from those of the previous MPL frameworks (e.g. \texttt{SecureML}~\cite{mohassel2017secureml} and \texttt{ABY3}~\cite{mohassel2018aby3}).

\vspace{-2mm}
\subsection{Robust Training}
\label{sec.robust}

The robustness employed in \pmpl ensures that training would continue even though one \textit{assistant party} drops out.
In \pmpl, an additional \textit{public vector}, referred to as the alternate vector, is held by the \textit{privileged party}. The alternate vector can be represented linearly by the vectors held by two \textit{assistant parties}.
Here, we denote all shares generated by the alternate vector as  alternate shares.
During training, if no \textit{assistant party} drops out, these alternate shares are executed with the same operations as other shares. Once one \textit{assistant party} drops out, the alternate shares would replace the shares held by the dropped party. Thus the rest two parties can continue training.

With the robustness, the \textit{privileged party} can tolerate the dropping out of one \textit{assistant party}, even though the \textit{assistant party} intentionally quit the training process. Furthermore, the \textit{privileged party} can choose one \textit{assistant party} to reveal the final model, thus keeping its privileged position until the end of the training.

\vspace{-2mm}
\section{Design of \pmpl}
\label{sec.design}
In this section, we firstly introduce the sharing semantics of \pmpl, as well as sharing and reconstruction protocols. After that, we show the basic primitives and the building blocks that are designed to support 3PC training in \pmpl. Furthermore, we introduce the design of robustness of \pmpl. Finally, we analyze the complexity of our proposed protocols.

\vspace{-2mm}
\subsection{Sharing Semantics}
\label{sec.ss}
In this paper, we leverage two types of secret sharing protocols,  $\langle \cdot \rangle$-sharing and $[ \cdot ]$-sharing:

\begin{itemize}[leftmargin=*]
\item $\langle \cdot \rangle$-sharing:
We use $\langle \cdot \rangle$ to denote the shares of vector space secret sharing.
The more detailed descriptions of sharing protocol and reconstruction protocol are shown in Section~\ref{sec:share_reveal}.

\item $[ \cdot ]$-sharing: We use $[ \cdot ]$ to denote the shares of additive secret sharing.
A value $x \in \mathbb{Z}_{2^\ell}$ is said to be $[ \cdot ]$-shared among a set of parties $\mathcal{P} = \{P_0, P_1, P_2 \}$, if each party $P_i$ holds $[ x ]_i \in \mathbb{Z}_{2^\ell}$ $ (i \in \{0,1,2 \})$, such that $x = ([ x ]_0 + [ x ]_1 +[ x ]_2) ~mod~2^\ell$, which is represented as $x = [ x ]_0 + [ x ]_1 +[ x ]_2$ in the rest of the paper.
Besides, we define the boolean sharing as $[ \cdot ]^2$, which refers to the shares over $\mathbb{Z}_{2}$.

\end{itemize}

Note that we use $\langle \cdot \rangle$-sharing as the underlying technique of \pmpl. Besides, $[\cdot]$-sharing is only used for the comparison protocol to represent the intermediate computation results.

\noindent\textbf{Linearity of the Secret Sharing Schemes:} Given the $\langle \cdot \rangle$-sharing of $x, y$ and public constants $k_1, k_2$, each party can locally compute $\langle k_1 \cdot x+k_2 \cdot y \rangle = k_1 \cdot \langle x \rangle + k_2 \cdot \langle y \rangle$. Besides, it is obvious that $[ \cdot ]$-sharing also satisfies the linearity property. The linearity property enables parties to non-interactively execute addition operations, as well as execute multiplication operations of their shares with a public constant.

\vspace{-2mm}
\subsection{Sharing and Reconstruction Protocols}
\label{sec:share_reveal}

In \pmpl, to share a secret value $x$, we form it as a three-dimensional vector $\vec{s} = (x, s_1, s_2)^T$, where $s_1$ and $s_2$ are two random values. 
We define a \textit{public matrix} $\varPhi (\mathcal{P})$ as a $4 \times 3$ matrix. 
Here, for each party $P_i$, the $i$-$\mathit{th}$ row $\varPhi(i)$ of $\varPhi (\mathcal{P})$ is its corresponding three-dimensional \textit{public vector}. 
Besides, the \textit{privileged party} $P_0$ holds the alternate three-dimensional \textit{public vector} $\varPhi(3)$.

To meet the two requirements mentioned in Section~\ref{sec.scenario}, the \textit{public matrix} $\varPhi (\mathcal{P})$ should satisfy four restrictions as follows:

\begin{itemize}[leftmargin=*]
    \item $(1,0,0)$ can be written as a linear combination of the \textit{public vectors} in the set $\{\varPhi(0), \varPhi(1), \varPhi(2)\}$, where $\varPhi(0), \varPhi(1), \varPhi(2)$ are linearly independent.
    Thus there are three non-zero public constants $c_0,c_1,c_2 $, such that $(1,0,0) = c_0 \cdot \varPhi(0) + c_1 \cdot \varPhi(1) + c_2 \cdot \varPhi(2)$.
    
    \item The \textit{public vector} $\varPhi(3)$ can be represented linearly by the vectors $\varPhi(1)$ and $\varPhi(2)$, i.e. $\varPhi(3) = a_1 \cdot \varPhi(1) + a_2 \cdot \varPhi(2)$, where $a_1, a_2 \neq 0$. Therefore, $(1,0,0)$ can also be written as a linear combination of the \textit{public vectors } in both sets $ \{ \varPhi(0), \varPhi(1), \varPhi(3) \}$ and $ \{ \varPhi(0), \varPhi(2), \varPhi(3)\}$.
    That is, there are six non-zero public constants $c'_0, c'_1, c'_3, c''_0, c''_2, c''_3 $, such that $(1,0,0) = c'_0 \cdot \varPhi(0) + c'_1 \cdot \varPhi(1) + c'_3 \cdot \varPhi(3) = c''_0 \cdot \varPhi(0) + c''_2 \cdot \varPhi(2) + c''_3 \cdot \varPhi(3)$.
    
    \item To prevent the set of parties that are not in the access structure from revealing the secret value, $(1,0,0)$ cannot be written as a linear combination of the \textit{public vectors} in both the sets $\{ \varPhi(0), \varPhi(3) \}$ and $\{\varPhi(1),\varPhi(2) \}$.
    
    \item As \pmpl performs the computations on the ring $\mathbb{Z}_{2^\ell}$, both the values of \textit{public matrix} $\varPhi (\mathcal{P})$ and reconstruction coefficients $c_0, c_1, \dots, c''_3$ should be elements of the ring $\mathbb{Z}_{2^\ell}$.
\end{itemize}

We formalize the above restrictions as Equation (\ref{eq.coefficient}) as follows:

\vspace{-2mm}
\begin{equation}
\label{eq.coefficient}
    \begin{aligned}
        (1,0,0) &= c_0 \cdot \varPhi(0) + c_1 \cdot \varPhi(1) + c_2 \cdot \varPhi(2)\\ 
        &= c'_0 \cdot \varPhi(0) + c'_1 \cdot \varPhi(1) + c'_3 \cdot \varPhi(3)\\
        &= c''_0 \cdot \varPhi(0) + c''_2 \cdot \varPhi(2) + c''_3 \cdot \varPhi(3)
    \end{aligned}
\end{equation}

Once the \textit{public matrix} $\varPhi (\mathcal{P})$ is determined, the reconstruction coefficients $c_0, c_1, \dots, c''_3$ can be computed by Equation (\ref{eq.coefficient}). It is trivial that these coefficients are also public to all parties.

\begin{algorithm}
    \LinesNumbered
    \small
    \caption{$\prod_{\rm shr} (P_i,x)$}
    \label{pro:ss}
    \begin{flushleft}
    \textbf{Input:}  The secret value $x$ held by $P_i$\\
    \textbf{Output:}  $\langle x \rangle$
    \end{flushleft}
    \begin{algorithmic}[1]
    \STATE $P_i$ constructs a three-dimensional vector $\vec{s} = (x, s_1, s_2)^T$, where $s_1$ and $s_2$ are random values.
    \STATE - If $P_i = P_0$, $P_i$ sends ${\langle x \rangle_j} =  \varPhi(j) \times \vec{s}$ to $P_j$ for $j \in \{ 1,2\}$. Meanwhile, $P_i$ generates ${\langle x \rangle_0} = \varPhi(0) \times \vec{s}$ and ${\langle x \rangle_3} = \varPhi(3) \times \vec{s}$ for itself.
    
    - If $P_i \neq P_0$, $P_i$ sends ${\langle x \rangle_j} =  \varPhi(j) \times \vec{s}$ to $P_j$ for $j \in \{ 0,1,2 \} \backslash \{i \}$, and sends the alternate share ${\langle x \rangle_3} = \varPhi(3) \times \vec{s}$ to $P_0$. Meanwhile, $P_i$ generates share ${\langle x \rangle_i} = \varPhi(i) \times \vec{s}$~for itself.
 \end{algorithmic}
\end{algorithm}
\vspace{-2mm}

\noindent\textbf{Sharing Protocol:} As is shown in Protocol~\ref{pro:ss}, $\prod_{\rm shr} (P_i,x)$ enables $P_i$ who holds the secret value $x$ to generate $\langle \cdot \rangle$-shares of $x$. In Step 1 of $\prod_{\rm shr} (P_i,x)$ (Protocol~\ref{pro:ss}), $P_i$ samples two random values $s_1$ and $s_2$ to construct a three-dimensional vector $\vec{s}$ $= (x, s_1, s_2)^T$. 
In Step 2 of $\prod_{\rm shr} (P_i,x)$ (Protocol \ref{pro:ss}), we consider two cases as follows: (1) If $P_i = P_0$, $P_i$ sends ${\langle x \rangle_j} = \varPhi(j) \times \vec{s}$ to two \textit{assistant parties} $P_j$ for $j \in \{ 1,2 \}$. Meanwhile, $P_i$ generates ${\langle x \rangle_0} = \varPhi(0) \times \vec{s}$ as well as the alternate share ${\langle x \rangle_3} = \varPhi(3) \times \vec{s}$, and holds them. 
(2) If $P_i \neq P_0$, $P_i$ sends ${\langle x \rangle_j} = \varPhi(j) \times \vec{s}$ to $P_j$ for $j \in \{ 0,1,2 \} \backslash \{i \}$. Besides, $P_i$ sends the alternate share ${\langle x \rangle_3} = \varPhi(3) \times \vec{s}$ to $P_0$ and holds ${\langle x \rangle_i} = \varPhi(i) \times \vec{s}$.
After the execution of $\prod_{\rm shr} (P_i,x)$ (Protocol \ref{pro:ss}), $P_0$ holds $\langle x \rangle_0$ and $\langle x \rangle_3$, $P_1$ holds $\langle x \rangle_1$, and $P_2$ holds $\langle x \rangle_2$. We use the standard real/ideal world paradigm to prove the security of $\prod_{\rm shr} (P_i,x)$ in Appendix~\ref{sec.proofs}.

\noindent\textbf{Reconstruction Protocol:}
According to Equation (\ref{eq.coefficient}) and $\prod_{\rm shr} (P_i,x)$ (Protocol \ref{pro:ss}), we can reveal the secret value $x$ through Equation (\ref{eq.rec_1}), (\ref{eq.rec_2}), or (\ref{eq.rec_3}) for different scenarios:

\vspace{-3mm}
\begin{align}
    x &= c_0\cdot \langle x \rangle_0 + c_1 \cdot \langle x \rangle_1 + c_2 \cdot \langle x \rangle_2 \label{eq.rec_1}\\ 
    &= c'_0 \cdot \langle x \rangle_0 + c'_1 \cdot \langle x \rangle_1 + c'_3 \cdot \langle x \rangle_3 \label{eq.rec_2}\\
    &=c''_0 \cdot \langle x \rangle_0 + c''_2 \cdot \langle x \rangle_2 + c''_3 \cdot \langle x \rangle_3 \label{eq.rec_3}
\end{align}

As is shown in Protocol \ref{pro:rec}, $\prod_{\rm rec} (\mathcal{P},\langle x \rangle )$ enables parties to reveal the secret value $x$. Without loss of generality, we assign $P_2$ as the dropping \textit{assistant party} when one party drops out, as is shown in Figure~\ref{fig.architecture}. We consider two cases as follows: (1) If no \textit{assistant party} drops out, each party $P_i$ receives shares from the other two parties. Then they compute Equation (\ref{eq.rec_1}) to reveal the secret value $x$ ( $P_i$ can also reveal the secret value $x$ by computing Equation (\ref{eq.rec_2}) or (\ref{eq.rec_3}).). 
(2) If $P_2$ drops out, $P_0$ receives the shares $\langle x \rangle_1$ from $P_1$. Meanwhile, $P_1$ receives the share $\langle x \rangle_0$ and $\langle x \rangle_3$ from $P_0$. Then $P_0$ and $P_1$ non-interactively compute Equation (\ref{eq.rec_2}) to reveal the secret value $x$ locally. Note that even though $P_1$ and $P_2$ collude with each other, without the participation of $P_0$, the secret value $x$ cannot be revealed in $\prod_{\rm rec} (\mathcal{P},\langle x \rangle )$ (Protocol \ref{pro:rec}). Besides, once training is completed, $P_1$ and $P_2$ send their shares to $P_0$, while $P_0$ does not send its final shares to other parties. Therefore, only $P_0$ can obtain the final model. Besides, we use the standard real/ideal world paradigm to prove the security of $\prod_{\rm rec} (\mathcal{P},\langle x \rangle )$ in Appendix~\ref{sec.proofs}.

\begin{algorithm}
    \LinesNumbered
    \small
    \caption{$\prod_{\rm rec} (\mathcal{P},\langle x \rangle )$}
    \label{pro:rec}
    \begin{flushleft}
    \textbf{Input:}  $\langle x \rangle$\\
    \textbf{Output:}  $x$ \\
    - If no party drops out: 
    \end{flushleft}
    \begin{algorithmic}[1]
        \STATE \quad $P_i$ receives shares from the other two parties.
    
        \STATE \quad$P_i$ reveal $x$ by computing Equations (\ref{eq.rec_1}): $x = c_0\cdot \langle x \rangle_0 + c_1 \cdot \langle x \rangle_1 + c_2 \cdot \langle x \rangle_2$.  
    \end{algorithmic}
    \begin{flushleft}
    - If $P_2$ drops out:
    \end{flushleft}
    \begin{algorithmic}[1]
        \STATE \quad$P_0$ receives $\langle x \rangle_1$ from $P_1$. Meanwhile, $P_1$ receives $\langle x \rangle_0$ and $\langle x \rangle_3$ from $P_0$.
    
        \STATE \quad$P_0$ and $P_1$ reveal $x$ by computing Equations (\ref{eq.rec_2}): $x=c'_0 \cdot \langle x \rangle_0 + c'_1 \cdot \langle x \rangle_1 + c'_3 \cdot \langle x \rangle_3$.
    \end{algorithmic}
\end{algorithm}
\vspace{-2mm}

\vspace{-2mm}
\subsection{Basic Primitives for 3PC}
\label{sec.basic}

In this section, we introduce the design of the basic primitives  in \pmpl for 3PC (i.e. no party drops out) in detail, including: (1) the primitives of secure addition and secure multiplication; (2) the primitives of sharing conversion: $\langle \cdot \rangle$-sharing to $[ \cdot ]$-sharing and $[ \cdot ]$-sharing to $\langle \cdot \rangle$-sharing; (3) MSB extraction and Bit2A, i.e. boolean to additive conversion. Besides, we use the standard real/ideal world paradigm to prove the security of these basic primitives in Appendix~\ref{sec.proofs}.

\noindent\textbf{Secure Addition:}
Given two secret values $x$ and $y$, each party $P_i$ holds shares ${\langle x \rangle_i}$ and ${\langle y \rangle_i}$ ($P_0$ additionally holds the alternate shares $\langle x \rangle_3$ and $\langle y \rangle_3$).
To get the result of secure addition $\langle x + y \rangle$, each party $P_i$ can utilize the linearity property of the $\langle \cdot \rangle$-sharing scheme to locally compute ${\langle z \rangle_i} = {\langle x \rangle_i} + {\langle y \rangle_i}$. $P_0$ additionally computes ${\langle z \rangle_3} = {\langle x \rangle_3} + {\langle y \rangle_3}$ for the alternate shares.

\noindent\textbf{Secure Multiplication:}
Through interactive computing, parties securely multiply two shares $\langle x \rangle$ and $\langle y \rangle$.
According to Equation (\ref{eq.mul}), we utilize two random values $u$ and $v$ to mask the secret values $x$ and $y$. More specifically, we utilize a vector multiplication triplet $(u,v,h)$, which refers to the method of Beaver's multiplication triplet~\cite{beaver1991efficient}, to execute secure multiplication.

\vspace{-2mm}
\begin{equation}
\label{eq.mul}
\begin{aligned}
    x \cdot y &= x \cdot (y+v)- x \cdot v 
    = x \cdot (y+v)-v \cdot (x+u-u)\\
    & = x \cdot (y+v)-v \cdot (x+u)+v \cdot u\\
\end{aligned}
\end{equation}

Protocol~\ref{pro:mul} shows the secure multiplication protocol $\prod_{\rm mul} (\mathcal{P},\langle x \rangle, \langle y \rangle)$ proposed in \pmpl. 
Besides, the shares held by each party during the execution of secure multiplication, which consists of five steps, are shown in Appendix~\ref{sec.mulshare}, (concretely in Table~\ref{table.mulshare}).
In the offline phase of $\prod_{\rm mul} (\mathcal{P},\langle x \rangle, \langle y \rangle)$ (Protocol \ref{pro:mul}), we set $\vec{r} = (u, r_1, r_2 )^T$,  $\vec{q} =$ $(v, q_1, q_2)^T$ uniformly random three-dimensional vectors and $\vec{t}= (h , t_1, t_2 )^T = (u \cdot v , t_1, t_2 )^T$, where $t_1$, $t_2$ are uniformly random values. 
We assume that all the parties have already shared vector multiplication triplet ($\langle u \rangle$, $\langle v \rangle$, $\langle h \rangle$) in the offline phase. 
In the online phase of $\prod_{\rm mul} (\mathcal{P},\langle x \rangle, \langle y \rangle)$ (Protocol \ref{pro:mul}), firstly, each party $P_i$ locally computes $\langle e \rangle_i = \langle x \rangle_i + \langle u \rangle_i$ and $\langle f \rangle_i =\langle y \rangle_i + \langle v \rangle_i$. 
$P_0$ additionally computes the alternate shares $\langle e \rangle_3 = \langle x \rangle_3 + \langle u \rangle_3$ and $\langle f \rangle_3 =\langle y \rangle_3 + \langle v \rangle_3$ locally. 
To get $e$ and $f$, parties then interactively execute $\prod_{\rm rec} (\mathcal{P},\langle e \rangle )$ (Protocol \ref{pro:rec}) and $\prod_{\rm rec} (\mathcal{P},\langle f \rangle )$ (Protocol \ref{pro:rec}). 
Finally, each party $P_i$ locally computes $\langle z \rangle_i = \langle x \rangle_i \cdot f - \langle v \rangle_i \cdot e + \langle h \rangle_i$. Similarly, $P_0$ additionally computes the alternate share $\langle z \rangle_3 = \langle x \rangle_3 \cdot f - \langle v \rangle_3 \cdot e + \langle h \rangle_3$.

\vspace{-2mm}
\begin{algorithm}
    \LinesNumbered
    \small
    \caption{$\prod_{\rm mul} (\mathcal{P},\langle x \rangle, \langle y \rangle)$}
    \label{pro:mul}
    \begin{flushleft}
    \textbf{Preprocessing:} Parties pre-shared vector multiplication triplet ${\langle u \rangle, \langle v \rangle, \langle h \rangle}$ using $\prod_{\rm vmtgen} (\mathcal{P})$ (Protocol \ref{pro:vmt})\\
    \textbf{Input:}  $\langle x \rangle$ and $\langle y \rangle$\\
    \textbf{Output:} $\langle x \cdot y \rangle$
    \end{flushleft}
    \begin{algorithmic}[1]
        \STATE $P_i$ locally computes $\langle e \rangle_i = \langle x \rangle_i + \langle u \rangle_i$ and $\langle f \rangle_i =\langle y \rangle_i + \langle v \rangle_i$. $P_0$ additionally computes $\langle e \rangle_3 = \langle x \rangle_3 + \langle u \rangle_3$ and $\langle f \rangle_3 =\langle y \rangle_3 + \langle v \rangle_3$. 
    
        \STATE Parties interactively execute $\prod_{\rm rec} (\mathcal{P},\langle e \rangle)$ (Protocol \ref{pro:rec}) and $\prod_{\rm rec} (\mathcal{P},\langle f \rangle )$ (Protocol \ref{pro:rec}).
    
        \STATE $P_i$ locally computes $\langle z \rangle_i = \langle x \rangle_i \cdot f - \langle v \rangle_i \cdot e+\langle h \rangle_i$ and $P_0$ additionally computes the alternate share $\langle z \rangle_3 = \langle x \rangle_3 \cdot f - \langle v \rangle_3 \cdot e+\langle h \rangle_3$. 
    \end{algorithmic}
\end{algorithm}
\vspace{-2mm}

The vector multiplication triplets can be generated by a cryptography service provider (CSP) or securely generated by multi-party collaboration. 
  $\prod_{\rm vmtgen} (\mathcal{P})$ (Protocol \ref{pro:vmt}) enables parties to securely generate expected shared vector multiplication triplets $(\langle u \rangle, \langle v \rangle, \langle h \rangle)$. It consists of two phases, i.e. generating ${\langle u \rangle, \langle v \rangle}$ and generating ${\langle h \rangle}$. Moreover, the shares that each party holds during the execution of $\prod_{\rm vmtgen} (\mathcal{P})$ (Protocol \ref{pro:vmt}), which consists of seven steps, are shown in Appendix~\ref{sec.vmtshare} (concretely in Table~\ref{table.vmtshare}).

\begin{itemize}[leftmargin=*]
    \item \emph{Generating $\langle u \rangle$ and $\langle v \rangle$:}
    As $\langle u \rangle$ and $\langle v \rangle$ are generated in the same way, we hereby take the generation of $\langle u \rangle$ as an example.
    Firstly, each party $P_i$ generates a random value $u_i$. Then they interactively execute $\prod_{\rm shr} (P_i,u_i)$ (Protocol \ref{pro:ss}). After that, each party $P_i$ holds three shares $\langle u_0 \rangle_i, \langle u_1 \rangle_i, \langle u_2 \rangle_i$. Besides, $P_0$ additionally holds another three alternate shares $\langle u_0 \rangle_3, \langle u_1 \rangle_3, \langle u_2 \rangle_3$. Then each party $P_i$ adds up these three shares locally to compute $\langle u \rangle_i = \langle u_0 \rangle_i +\langle u_1 \rangle_i + \langle u_2 \rangle_i$. $P_0$ additionally computes  $\langle u \rangle_3 = \langle u_0 \rangle_3 +\langle u_1 \rangle_3 + \langle u_2 \rangle_3$.    
\end{itemize}

\begin{itemize}[leftmargin=*]
    \item \emph{Generating $\langle h \rangle$:} Given shared random values $\langle u \rangle$ and $\langle v \rangle$ mentioned above, the key step of generating $\langle h \rangle$ is to compute the shares of their product. According to the process of generating $\langle u \rangle$ and $\langle v \rangle$, we can get that $u = u_0+u_1+u_2$ and $v = v_0+v_1+v_2$. Then:
    
   \begin{equation}
    \begin{aligned}
        h = u v = (u_0+u_1+u_2)(v_0+v_1+v_2) 
        = u_0 v_0 +u_0 v_1+u_0 v_2 \\
        + u_1 v_0 +u_1 v_1+u_1 v_2 
        + u_2 v_0 +u_2 v_1+u_2 v_2
    \end{aligned}
    \end{equation}
    where $u_i  v_i$ ($i\in \{0, 1, 2\}$)  can be computed locally in each party $P_i$ and the rest products require three parties to compute cooperatively. We use the method proposed by Zhu and Takagi~\cite{zhu2015efficient} to calculate  $[ u_0 v_1+u_1 v_0 ]$, $[ u_0 v_2+u_2 v_0 ]$, and $[ u_1 v_2+u_2 v_1 ]$.  After that, each party $P_i$ locally computes
    $h_i = u_i v_i+ [ u_i v_{i+1}+ u_{i+1} v_i]_i+[ u_i v_{i-1}+u_{i-1} v_i]_i$. Here, $i \pm 1$ refers to the next (+) or previous (-) party with wrap around. For example, the party 2 + 1 is the party 0, and the party 0 - 1 is the party 2. 
    Subsequently, each party $P_i$ executes $\prod_{\rm shr} (P_i,h_i)$ (Protocol \ref{pro:ss}) to get three shares $\langle h_0 \rangle_i, \langle h_1 \rangle_i$ and $\langle h_2 \rangle_i$ ($P_0$ additionally holds three alternate shares $\langle h_0 \rangle_3, \langle h_1 \rangle_3$ and $\langle h_2 \rangle_3$). At last, each party $P_i$ adds up the three shares locally to get $\langle h \rangle_i = \langle h_0 \rangle_i + \langle h_1 \rangle_i + \langle h_2 \rangle_i$ ($P_0$ additionally adds up three alternate shares to get $\langle h \rangle_3 = \langle h_0 \rangle_3 + \langle h_1 \rangle_3 + \langle h_2 \rangle_3$).
\end{itemize}

\vspace{-2mm}
\begin{algorithm}
    \LinesNumbered
    \small
    \caption{$\prod_{\rm vmtgen} (\mathcal{P})$}
    \label{pro:vmt}
    \begin{flushleft}
    \textbf{Input:} $\emptyset$\\
    \textbf{Output:}  The shares of vector multiplication triplet $({\langle u \rangle, \langle v \rangle, \langle h \rangle})$
    \textbf{Generating ${\langle u \rangle, \langle v \rangle}$:}
    \end{flushleft}
    \begin{algorithmic}[1]
        \STATE $P_i$ generates two random values $u_i$ and $v_{i}$.
    
        \STATE $P_i$ executes $\prod_{\rm shr} (P_i,u_i)$ (Protocol~\ref{pro:ss}) and $\prod_{\rm shr} (P_i,v_i)$ (Protocol \ref{pro:ss}).
    
        \STATE $P_i$ locally computes $\langle u \rangle_i = \langle u_0 \rangle_i +\langle u_1 \rangle_i+\langle u_2 \rangle_i$, and $\langle v \rangle_i = \langle v_0 \rangle_i +\langle v_1 \rangle_i+\langle v_2 \rangle_i$. Besides, $P_0$ computes the alternate shares $\langle u \rangle_3$ and $\langle v \rangle_3$ in the same way. 
    \end{algorithmic}

    \begin{flushleft}
    \textbf{Generating ${\langle h \rangle}$:}
    \end{flushleft}
    \begin{algorithmic}[1]
        \STATE $P_0$ and $P_1$ interactively compute $[ u_0 v_1+u_1 v_0 ]$, $P_0$ and $P_2$ interactively compute $[ u_0 v_2+u_2 v_0 ]$, $P_1$ and $P_2$ interactively compute $[ u_1 v_2+u_2 v_1 ]$.
    
        \STATE $P_i$ locally computes $h_i = u_i v_i+ [ u_i v_{i+1}+ u_{i+1} v_i]_i+[ u_i v_{i-1}+u_{i-1} v_i]_i$. 
    
        \STATE $P_i$ executes $\prod_{\rm shr} (P_i,h_i)$ (Protocol \ref{pro:ss}).
    
        \STATE $P_i$ locally computes $\langle h \rangle_i = \langle h_0 \rangle_i +\langle h_1 \rangle_i+\langle h_2 \rangle_i$ and $P_0$ additionally computes the alternate share $\langle h \rangle_3 = \langle h_0 \rangle_3 +\langle h_1 \rangle_3+\langle h_2 \rangle_3$. 
    
    \end{algorithmic}
\end{algorithm}
\vspace{-2mm}

\noindent\textbf{Sharing Conversion:}
Previous studies~\cite{koti2021swift}\cite{mohassel2018aby3} have established that non-linear operations such as comparison are more efficient in $\mathbb{Z}_{2}$ than in  $\mathbb{Z}_{2^\ell}$. That is, $[ \cdot ]^2$-sharing is more suitable for executing non-linear operations than both $ \langle \cdot \rangle$-sharing and $[ \cdot ]$-sharing.
However, the conversions between $ \langle \cdot \rangle$-shares and $[ \cdot ]^2$-shares are challenging, while the conversions between $\langle \cdot \rangle$-shares and $[ \cdot ]$-shares are relatively easy to perform. 
Thus, to efficiently execute non-linear operations, we firstly convert $ \langle \cdot \rangle$-shares to
$[ \cdot ]$-shares locally. Furthermore, we use the existing methods~\cite{DBLP:conf/sp/Damgard0FKSV19}\cite{mohassel2018aby3} to convert between $ [\cdot ]$-shares and $[ \cdot ]^2$-shares. 
Finally, we convert $[ \cdot ]$-shares back to $ \langle \cdot \rangle$-shares.

We hereby present two primitives of sharing conversion as follows:

\begin{itemize}[leftmargin=*]
    \item \emph{Converting $\langle \cdot \rangle${-shares} to $[ \cdot ]${-shares}:} 
     $\prod_{\rm v2a} (\mathcal{P},\langle x \rangle )$ enables
    each party $P_i$ locally computes $[x]_i=c_i \cdot \langle x \rangle_i$ to convert $\langle \cdot \rangle$-shares to $[ \cdot ]$-shares according to Equation (\ref{eq.conv}).

    \vspace{-4mm}
    \begin{equation}
    \label{eq.conv}
        \begin{aligned}
        x = c_0\cdot \langle x \rangle_0 + c_1 \cdot \langle x \rangle_1 + c_2 \cdot \langle x \rangle_2 = [x]_0+[x]_1+[x]_2
        \end{aligned}
    \end{equation}

    Here, we only convert three, i.e. $\langle x \rangle_0$, $\langle x \rangle_1$, $\langle x \rangle_2$, of the four $\langle \cdot \rangle$-shares to $[ \cdot ]$-shares. 
    Since \pmpl supports the privileged party and one   of two assistant parties (three shares) to train and the reconstruction protocol only needs three shares, this configuration does not affect subsequent operations.

    \item \emph{Converting $[ \cdot ]${-shares} to $\langle \cdot \rangle${-shares}:}
    $\prod_{\rm a2v} (\mathcal{P},[ x ] )$ (Protocol \ref{pro:a2v}) enables parties to convert $[ \cdot ]$-sharing to $\langle \cdot \rangle$-sharing.
    Here, we are supposed to convert three $[\cdot]$-shares to four $\langle \cdot \rangle$-shares. 
    Except for the alternate share, each party $P_i$ locally computes $\langle x \rangle_i = [x]_i /c_i$. 
    Due to the equation: $\varPhi(3) = a_1 \cdot \varPhi(1) + a_2 \cdot \varPhi(2)$, we can get the alternate share $\langle x \rangle_3$ by computing $\langle x \rangle_3 = a_1 \cdot \langle x \rangle_1 + a_2 \cdot \langle x \rangle_2$. 
    We assume that all the parties have already shared a random value $k$, which is generated in the same way as $\langle u \rangle$ and $\langle v \rangle$ in $\prod_{\rm vmtgen} (\mathcal{P})$ (Protocol \ref{pro:vmt}).
    Then $P_1$ and $P_2$ compute $\langle x \rangle_j + \langle k \rangle_j$ ($j \in \{1,2\}$) locally, and send them in plaintext to $P_0$. Finally, $P_0$ locally computes the alternate share $\langle x \rangle_3 = a_1 \cdot (\langle x \rangle_1 + \langle k \rangle_1)+a_2 \cdot (\langle x \rangle_2 + \langle k \rangle_2) - \langle k \rangle_3$.

\end{itemize}

\begin{algorithm}
    \LinesNumbered
    \small
    \caption{$\prod_{\rm a2v} (\mathcal{P},[ x ] )$}
    \label{pro:a2v}
    \begin{flushleft}
    \textbf{Preprocessing:} Parties pre-shared $\langle k \rangle$ \\ 
    \textbf{Input:}  $[ x ]$\\
    \textbf{Output:}  $\langle x \rangle$
    \end{flushleft}
    \begin{algorithmic}[1]
        \STATE $P_i$ locally computes $\langle x \rangle_i = [x]_i /c_i$.
    
        \STATE $P_1$ and $P_2$ locally compute $\langle x \rangle_j + \langle k \rangle_j$ ($j \in \{1,2\}$) , and send them to $P_0$.
    
        \STATE $P_0$ locally computes $\langle x \rangle_3 = a_1 \cdot (\langle x \rangle_1 + \langle k \rangle_1)+a_2 \cdot (\langle x \rangle_2 + \langle k \rangle_2) - \langle k \rangle_3$.
    \end{algorithmic}
\end{algorithm}

\noindent\textbf{MSB extraction and Bit2A:}
The MSB extraction protocol $\prod_{\rm msbext} (\mathcal{P}, [ x ])$  enables parties to compute boolean sharing of MSB of a value $x$ (Here, we use the method presented in the study~\cite{DBLP:conf/fc/MakriRVW21}, and name it in this paper). 
Bit2A protocol $\prod_{\rm b2a} (\mathcal{P}, [ b ]^2)$ enables parties to compute from the boolean sharing of $b$ ($[b]^2$) to its additive secret sharing ($[b]$) (Here, we use the method presented in the study~\cite{DBLP:conf/sp/Damgard0FKSV19}, and name it in this paper).

\subsection{Building Blocks for \pmpl}

We detail the design of the building blocks in \pmpl for 3PC as follows: (1) matrix sharing; (2) matrix addition and matrix multiplication; (3) truncation; (4) two activation functions, i.e. \texttt{ReLU} and \texttt{Sigmoid}.

\begin{figure}[h]
\centering
\vspace{-3mm}
\includegraphics[scale=0.5]{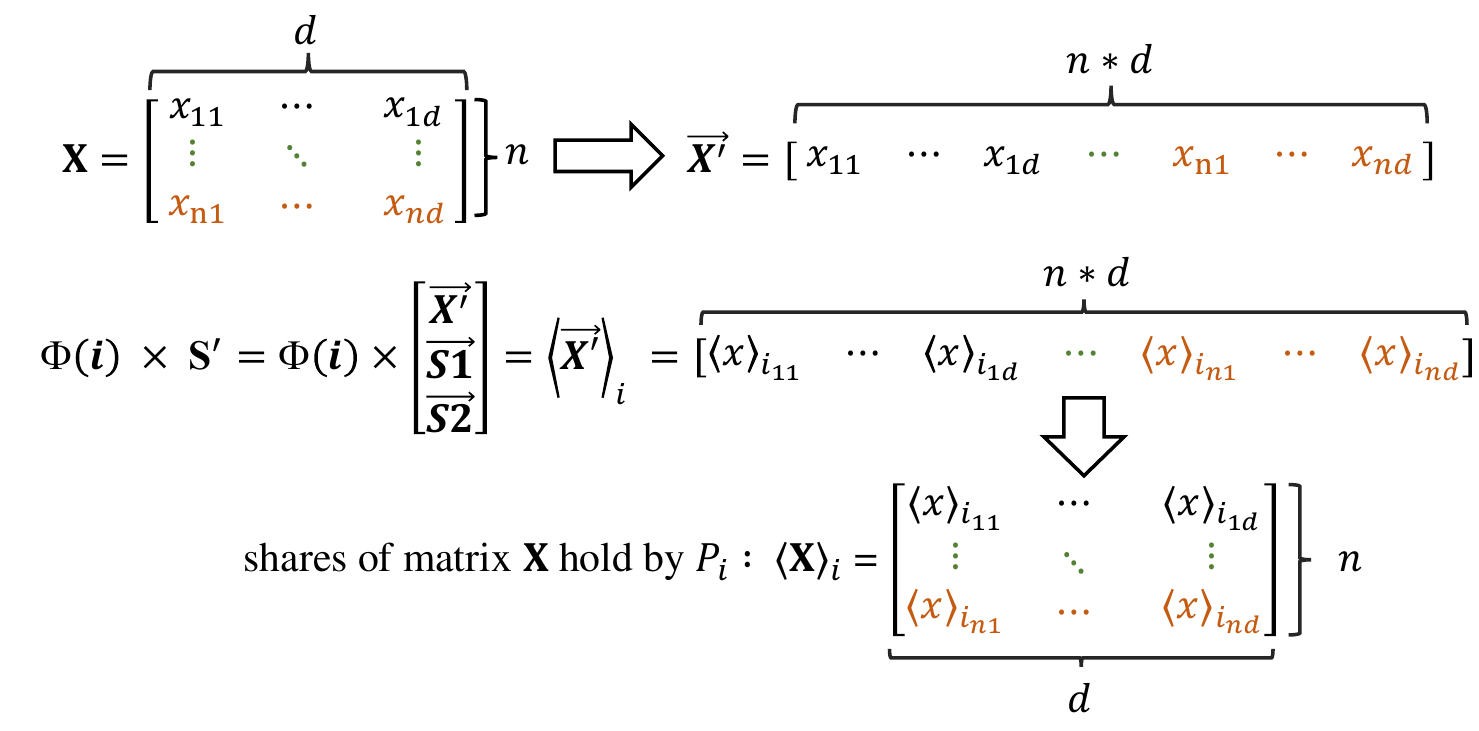}
\vspace{-3mm}
\caption{Matrix conversions during matrix sharing}
\vspace{-3mm}
\label{fig.matrix_share}
\end{figure}

\noindent\textbf{Matrix Sharing:}
As all the variables in \pmpl are represented as matrices. In order to improve the efficiency of sharing protocol, we generalize the sharing operation on a single secret value to an $n \times d$ secret matrix $\mathbf{X}$. As is shown in Figure \ref{fig.matrix_share}, $P_i$ who holds the secret matrix $\mathbf{X}$ firstly flattens $\mathbf{X}$ into row vector $\vec{X'}$ with the size of $nd$. Then $P_i$ constructs a $3 \times nd$ matrix $\mathbf{S'} =  (\vec{X'}^T,\vec{S1}^T,\vec{S2}^T)^T$, where $\vec{S1}$ and $\vec{S2}$ are random row vectors with size of $nd$.
Furthermore, $P_i$ computes shares $\langle \vec{X'} \rangle_k = \varPhi(k) \times \mathbf{S}'$ for $k = \{0,1,2,3\}$.
Finally, $P_i$ converts $\langle \vec{X'} \rangle_k$ to an $n \times d$ matrix $\langle \mathbf{X} \rangle_k$.

\noindent\textbf{Matrix Addition and Multiplication:}
We generalize the addition and multiplication operations on shares to shared matrices referring to the method of~\cite{mohassel2017secureml}. 
Given two shared matrices $\langle \mathbf{X} \rangle$ (with the size of $n \times d$) and $\langle \mathbf{Y} \rangle$ (with the size of $d \times m$), in the matrix addition, each party $P_i$ locally computes ${\langle \mathbf{Z} \rangle_i} = {\langle \mathbf{X} \rangle_i} + {\langle \mathbf{Y} \rangle_i}$. $P_0$ additionally computes the alternate shared matrix ${\langle \mathbf{Z} \rangle_3} = {\langle \mathbf{X} \rangle_3} + {\langle \mathbf{Y} \rangle_3}$.
To multiply two shared matrices $\langle \mathbf{X} \rangle$ and $\langle \mathbf{Y} \rangle$, instead of using independent vector multiplication triplets ($u,v,h$) on each element multiplication, we take matrix vector multiplication triplets ($\mathbf{U}, \mathbf{V}, \mathbf{H}$) to execute the matrix multiplication.
Here, $\mathbf{U}$ and $\mathbf{V}$ are random matrices, $\mathbf{U}$ has the same dimension as $\mathbf{X}$, $\mathbf{V}$ has the same dimension as $\mathbf{Y}$ and $\mathbf{H} = \mathbf{U} \times \mathbf{V}$. We assume that all the parties have already shared ($\langle \mathbf{U} \rangle$, $\langle \mathbf{V} \rangle$, $\langle \mathbf{H} \rangle$).  Each party $P_i$ firstly computes $\langle \mathbf{E} \rangle_i = \langle \mathbf{X} \rangle_i + \langle \mathbf{U} \rangle_i$ and $\langle \mathbf{F} \rangle_i =\langle \mathbf{Y} \rangle_i + \langle \mathbf{V} \rangle_i$ locally. $P_0$ additionally computes $\langle \mathbf{E} \rangle_3 = \langle \mathbf{X} \rangle_3 + \langle \mathbf{U} \rangle_3$ and $\langle \mathbf{F} \rangle_3 =\langle \mathbf{Y} \rangle_3 + \langle \mathbf{V} \rangle_3$.  Then parties reveal $\mathbf{E}$ and $\mathbf{F}$, and compute $\langle \mathbf{Z} \rangle_i = \langle \mathbf{X} \rangle_i \times \mathbf{F} - \mathbf{E} \times \langle \mathbf{V} \rangle_i + \langle \mathbf{H} \rangle_i$ locally. $P_0$ additionally computes $\langle \mathbf{Z} \rangle_3 = \langle \mathbf{X} \rangle_3 \times \mathbf{F} - \mathbf{E} \times \langle \mathbf{V} \rangle_3 + \langle \mathbf{H} \rangle_3$.

As for the generation of matrix vector multiplication triplets ($\mathbf{U}, \mathbf{V}, \mathbf{H}$), the process is similar to $\prod_{\rm vmtgen} (\mathcal{P})$ (Protocol \ref{pro:vmt}), where the sharing protocol is replaced with the matrix sharing protocol.
For the generation of $\mathbf{U}$ and $\mathbf{V}$, we also take $\mathbf{U}$ as an example.
Firstly, each party $P_i$ generates a random $n \times d$ matrix   $\mathbf{U}_i$, $P_3$ additionally generates a random matrix $\mathbf{U}_3$. 
Then each party $P_i$ shares (using matrix sharing protocol) $\mathbf{U}_i$, $P_3$ additionally shares matrices $\mathbf{U}_3$. 
After that, each party $P_i$ holds three shared matrices $\langle \mathbf{U}_0 \rangle_i, \langle \mathbf{U}_1 \rangle_i, \langle \mathbf{U}_2 \rangle_i$. Besides, $P_0$ additionally holds another three alternate shares $\langle \mathbf{U}_0 \rangle_3, \langle \mathbf{U}_1 \rangle_3, \langle \mathbf{U}_2 \rangle_3$. Then each party $P_i$ adds these three shared matrices locally to compute $\langle \mathbf{U} \rangle_i = \langle \mathbf{U}_0 \rangle_i +\langle \mathbf{U}_1 \rangle_i + \langle \mathbf{U}_2 \rangle_i$. Additionally, $P_0$ computes $\langle \mathbf{U} \rangle_3 = \langle \mathbf{U}_0 \rangle_3 +\langle \mathbf{U}_1 \rangle_3 + \langle \mathbf{U}_2 \rangle_3$.
For the generation of $\langle \mathbf{H} \rangle$, we generalize the secure computation method proposed by Zhu and Takagi~\cite{zhu2015efficient} to shared matrices. 
Firstly, $P_0$ and $P_1$ interactively compute $[ \mathbf{U}_0 \times \mathbf{V}_1+\mathbf{U}_1 \times \mathbf{V}_0 ]$, $P_0$ and $P_2$ interactively compute $[ \mathbf{U}_0 \times \mathbf{V}_2+\mathbf{U}_2 \times \mathbf{V}_0 ]$, $P_1$ and $P_2$ interactively compute $[ \mathbf{U}_1 \times \mathbf{V}_2+\mathbf{U}_2 \times \mathbf{V}_1 ]$. Then each party $P_i$ locally computes $\mathbf{H}_i = \mathbf{U}_i \times \mathbf{V}_i+ [ \mathbf{U}_i \times \mathbf{V}_{i+1}+ \mathbf{U}_{i+1} \times \mathbf{V}_i]_i+[ \mathbf{U}_i \times \mathbf{V}_{i-1}+\mathbf{U}_{i-1} \times \mathbf{V}_i]_i$. Furthermore, each party $P_i$ shares $\mathbf{H}_i$ using the matrix sharing protocol. Finally, each party $P_i$ locally computes $\langle \mathbf{H} \rangle_i = \langle \mathbf{H}_0 \rangle_i +\langle \mathbf{H}_1 \rangle_i+\langle \mathbf{H}_2 \rangle_i$. $P_0$ additionally computes the alternate shared matrix $\langle \mathbf{H} \rangle_3 = \langle \mathbf{H}_0 \rangle_3 +\langle \mathbf{H}_1 \rangle_3+\langle \mathbf{H}_2 \rangle_3$.

\vspace{-2mm}
\begin{algorithm}
    \LinesNumbered
    \small
    \caption{$\prod_{\rm trunc} (\mathcal{P}, \langle z \rangle)$}
    \label{pro:trunc}
    \begin{flushleft}
    \textbf{Preprocessing:} Parties pre-shared random values $\langle r \rangle$ and $\langle r' \rangle = \langle r / 2^{\ell_f} \rangle$ \\ 
    \textbf{Input:}  $\langle z \rangle$\\
    \textbf{Output:}  The result after truncation $\langle z' \rangle$, where $z' = z/ 2^{\ell_f}$
    \end{flushleft}
    \begin{algorithmic}[1]
        \STATE $P_i$ locally computes $\langle z - r \rangle_i = \langle z \rangle_i - \langle r \rangle_i$. $P_0$ additionally computes $\langle z - r \rangle_3 = \langle z \rangle_3 - \langle r \rangle_3$;
    
        \STATE $P_1$ and $P_2$ send $\langle z - r \rangle_1$ and $\langle z -r \rangle_2$ to $P_0$ respectively.
    
        \STATE $P_0$ locally computes  $ \langle z' \rangle_0 = (z-r)/(2^{\ell_f} \cdot c_0) + \langle r' \rangle_0$ and \textit{assistant parties} $P_j$ for $j\in \{1,2\}$ holds $\langle z \rangle_j = \langle r' \rangle_j$. $P_0$ additionally holds $\langle z' \rangle_3 = \langle r' \rangle_3$.
    \end{algorithmic}
\end{algorithm}
\vspace{-2mm}

\noindent\textbf{Truncation:}
After multiplying two fixed-point numbers with $\ell_f$ bits in the fractional part, the fractional part of the computation result is extended to $2\ell_f$ bits. In order to return the result of the multiplication back to the same format as that of the inputs,  parties interactively execute the truncation on the result of the multiplication.

Protocol \ref{pro:trunc} shows the truncation protocol $\prod_{\rm trunc} (\mathcal{P}, \langle z \rangle)$ proposed in \pmpl. At first, we observe that:

\vspace{-2mm}
\begin{equation}
\begin{aligned}
\label{eq.tru}
    z' &= \frac{z}{2^{\ell_f}} 
    = \frac{c_0 \cdot \langle z \rangle_0 + c_1 \cdot \langle z \rangle_1 + c_2 \cdot \langle z \rangle_2}{2^{\ell_f}} \\
    &= \frac{\parbox{6cm}{$c_0 \cdot (\langle z \rangle_0-\langle r \rangle_0+\langle r \rangle_0)+c_1 \cdot (\langle z \rangle_1-\langle r \rangle_1+\langle r \rangle_1) +c_2 \cdot (\langle z \rangle_2-\langle r \rangle_2+\langle r \rangle_2)$}}{2^{\ell_f}} \\
    &= \frac{(z-r)+c_0 \cdot \langle r \rangle_0+c_1 \cdot \langle r \rangle_1+c_2 \cdot \langle r \rangle_2}{2^{\ell_f}} \\
    &= \frac{z-r}{2^{\ell_f}} + c_0 \cdot \frac{\langle r \rangle_0}{2^{\ell_f}}+c_1 \cdot \frac{\langle r \rangle_1}{2^{\ell_f}} +c_2 \cdot \frac{\langle r \rangle_2}{2^{\ell_f}} \\
    &= c_0 \cdot \frac{(z-r)/c_0+\langle r \rangle_0}{2^{\ell_f}}+c_1 \cdot \frac{\langle r \rangle_1}{2^{\ell_f}} +c_2 \cdot \frac{\langle r \rangle_2}{2^{\ell_f}}
\end{aligned}
\end{equation}

We assume that parties have held the shares $\langle r \rangle$ and $\langle r' \rangle = \langle r / 2^{\ell_f} \rangle$. To compute the shares of $z' = z/2^{\ell_f} = (x \cdot y)/2^{\ell_f}$, $P_1$ and $P_2$ sends $\langle z - r \rangle_1$ and $\langle z - r \rangle_2$ to $P_0$ respectively. 
Then $P_0$ locally computes $ z-r = c_0 \cdot \langle z-r \rangle_0 + c_1 \cdot \langle z - r \rangle_1 + c_2 \cdot \langle z - r \rangle_2$,  $(z-r)/(2^{\ell_f} \cdot c_0) + \langle r' \rangle_0$, and $P_1$, $P_2$ hold $\langle r' \rangle_1$,$\langle r' \rangle_2$, respectively. Additionally, $P_0$ holds $\langle r' \rangle_3$.
Finally, the shares $\langle z \rangle$ are truncated.

For truncation pairs, we use some edabits~\cite{escudero2020improved} to generate them.
The edabits are used in the share conversation between $[\cdot]$ and $[\cdot]^2$. An edabit consists of a value $r$ in $\mathbb{Z}_{2^\ell}$, together with a set of $\ell$ random bits $(r_0, \dots, r_{\ell-1})$ shared in the boolean world, where $r=\sum_{i=0}^{\ell-1}2^i \cdot r_i$.  
$\prod_{\rm trunpair} (\mathcal{P})$ (Protocol \ref{pro:trun_pair}) shows how to generate truncation pairs.
Firstly, parties generate edabits $([r], [r_0]^2, [r_1]^2, \dots, [r_{\ell-1}]^2)$ and $([r'], [r'_0]^2, [r'_1]^2, \dots, [r'_{\ell-{\ell_f}-1}]^2)$, where $r' = r / 2 ^{\ell_f}$. 
After that, each party holds $[\cdot]$-sharing of $r$. Then they interactively execute $\prod_{\rm a2v} (\mathcal{P},[ r ] )$ and $\prod_{\rm a2v} (\mathcal{P},[ r' ] )$ (Protocol \ref{pro:a2v}) to get $\langle r \rangle$ and $\langle r' \rangle$.

\begin{algorithm}
    \LinesNumbered
    \small
    \caption{$\prod_{\rm trunpair} (\mathcal{P})$}
    \label{pro:trun_pair}
    \begin{flushleft}
    \textbf{Input:} $\emptyset$ \\
    \textbf{Output:}  The truncation pairs ($\langle r \rangle$, $\langle r' \rangle)$, where  $r'= r / 2^{\ell_f}$
    \end{flushleft}
    \begin{algorithmic}[1]
        \STATE Parties generate edabits $[r], [r_0]^2, [r_1]^2, \dots, [r_{\ell-1}]^2$ and $[r'], [r'_0]^2, [r'_1]^2, \dots, [r'_{\ell-{\ell_f}-1}]^2$.
    
        \STATE Parties interactively execute protocol $\prod_{\rm a2v} (\mathcal{P},[ r ] )$ and $\prod_{\rm a2v} (\mathcal{P},[ r' ] )$ (Protocol \ref{pro:a2v}).
    
    \end{algorithmic}
\end{algorithm}
\vspace{-2mm}

\noindent\textbf{Activation Functions:}  
We consider two widely used non-linear activation functions in machine learning, i.e. \texttt{ReLU} and \texttt{Sigmoid}. 
Besides, we describe the approximations and computations of these activation functions in \pmpl as follows.

\begin{algorithm}
    \LinesNumbered
    \small
    \caption{$\prod_{\rm relu} (\mathcal{P}, \langle x \rangle)$}
    \label{pro:relu}
    \begin{flushleft}
    \textbf{Input:} $\langle x \rangle$
 
    \textbf{Output:} $\langle \texttt{ReLU(x)} \rangle$, where $\texttt{ReLU(x)} = 0$ if $x < 0$ and $x$ otherwise
    \end{flushleft}
    \begin{algorithmic}[1]
        \STATE Parties locally execute $\prod_{\rm v2a} (\mathcal{P},\langle x \rangle )$ to obtain $[ x ]$.
    
        \STATE Parties interactively execute $\prod_{\rm msbext} (\mathcal{P},[ x ] )$ to obtain $[b]^2$.
        
        \STATE $P_i$ computes $[1 \oplus b]^2$ locally.
        
        \STATE Parties interactively execute $\prod_{\rm b2a} (\mathcal{P},[1 \oplus b]^2 )$ to obtain $ [1 \oplus b]$.
        
        \STATE Parties interactively execute $\prod_{\rm a2v}(\mathcal{P},[1 \oplus b])$ (Protocol \ref{pro:a2v}) to obtain $\langle 1 \oplus b \rangle$.
        
        \STATE Parties interactively execute $\prod_{\rm mul} (\mathcal{P},\langle 1 \oplus b \rangle, \langle x \rangle)$ (Protocol \ref{pro:mul}) to compute $\langle \texttt{ReLU(x)} \rangle$
    
    \end{algorithmic}
\end{algorithm}

\begin{itemize}[leftmargin=*]
    \item \emph{ReLU:} \texttt{ReLU} function, which is defined as $\texttt{ReLU(x)}=max(x,0)$, can be viewed as $\texttt{ReLU(x)}=(1 \oplus b)\cdot x$. The bit $b$ denotes the MSB of $x$, where $b=1$ if $x<0$ and 0 otherwise. 
    $\prod_{\rm relu} (\mathcal{P}, \langle x \rangle)$ (Protocol \ref{pro:relu}) enables parties to compute the shares of \texttt{ReLU} function outputs, $\langle \texttt{ReLU(x)} \rangle$.
    Firstly, parties interactively execute $\prod_{\rm v2a} (\mathcal{P},\langle x \rangle )$ to convert $\langle x \rangle$ to $[ x ]$. Then they interactively execute $\prod_{\rm msbext} (\mathcal{P},[ x ] )$ on $[x]$ to obtain the share of MSB of $x$, namely $[b]^2$. Furthermore, each party $P_i$ locally computes $[1 \oplus b]^2$. 
    Next, parties interactively execute $\prod_{\rm b2a} (\mathcal{P},[1 \oplus b]^2 )$ to convert $[1 \oplus b]^2$ to $ [1 \oplus b]$. After that, parties interactively execute $\prod_{\rm a2v}(\mathcal{P},[1 \oplus b])$ (Protocol \ref{pro:a2v}) to convert $ [1 \oplus b]$ to $\langle 1 \oplus b \rangle$. At last, parties interactively execute $\prod_{\rm mul} (\mathcal{P},\langle 1 \oplus b \rangle, \langle x \rangle)$ (Protocol \ref{pro:mul}) to compute $\langle \texttt{ReLU(x)} \rangle$, such that $\texttt{ReLU(x)} = 0$ if $x < 0$, and $\texttt{ReLU(x)} = x$ otherwise.

    \item \emph{Sigmoid:} \texttt{Sigmoid} function is defied as $\texttt{Sigmoid}(x) = 1/{(1+e^{-x})}$. In this paper, we use an MPC-friendly version~\cite{mohassel2017secureml} of the \texttt{Sigmoid} function, which is defined as: 

    \begin{equation}
    \label{eq.sigmoid}
        \texttt{Sigmoid}(x)=\left\{
    \begin{aligned}
    0 & , & x \le -\frac{1}{2} \\
    x+\frac{1}{2} & , & -\frac{1}{2} < x < \frac{1}{2} \\
    1 & , & x \ge \frac{1}{2} \\
    \end{aligned}
    \right.
    \end{equation}
    This function can be viewed as $\texttt{Sigmoid}(x) = (1 \oplus b_1) \cdot b_2 \cdot (x + 1/2) + (1 \oplus b_2)$, where  $b_1 = 1$ if $x < -1/2$ and $b_2 = 1$ if $x < 1/2$.
    $\prod_{\rm sig} (\mathcal{P}, \langle x \rangle)$ is similar to $\prod_{\rm relu} (\mathcal{P}, \langle x \rangle)$. We thus do not describe it in detail.

\end{itemize}

\vspace{-2mm}
\subsection{Robustness Design (2PC)}
In \pmpl, we ensure the robustness through the design of the alternate shares. If $P_2$ drops out, the alternate shares will replace the shares held by $P_2$. Therefore, even if one \textit{assistant party} ($P_2$) drops out, the remaining two parties ($P_0$ and $P_1$) can continue training.
Here, we describe the protocols for the scenario of one of two \textit{assistant parties} ($P_2$) drops out, i.e. 2PC protocols.

\noindent\textbf{Secure Addition and Secure Multiplication:} 
To get the result of secure addition $\langle x + y \rangle$, if $P_2$ drops out, $P_0$ locally computes $\langle z \rangle_0 = \langle x \rangle_0 + \langle y \rangle_0$, $\langle z \rangle_3 = \langle x \rangle_3 + \langle y \rangle_3$, and $P_1$ locally computes $\langle z \rangle_1 = \langle x \rangle_1 + \langle y \rangle_1$.

\begin{algorithm}
    \LinesNumbered
    \small
    \caption{$\prod_{\rm mul2} (\mathcal{P},\langle x \rangle,
\langle y \rangle)$}
    \label{pro:mul2}
    \begin{flushleft}
    \textbf{Preprocessing:} Parties pre-shared vector multiplication triplet ${\langle u \rangle, \langle v \rangle, \langle h \rangle}$ using $\prod_{\rm vmtgen} (\mathcal{P})$ (Protocol \ref{pro:vmt})  \\
    \textbf{Input:}  $\langle x \rangle$ and $\langle y \rangle$.\\
    \textbf{Output:} $\langle x \cdot y \rangle$.
    \end{flushleft}
    \begin{algorithmic}[1]
        \STATE $P_j$ for $j \in \{0,1\}$ locally computes $\langle e \rangle_j = \langle x \rangle_j + \langle u \rangle_j$ and $\langle f \rangle_j =\langle y \rangle_j + \langle v \rangle_j$. 
    Besides, $P_0$ computes $\langle e \rangle_3 = \langle x \rangle_3 + \langle u \rangle_3$ and $\langle f \rangle_3 =\langle y \rangle_3 + \langle v \rangle_3$. 
    
        \STATE Parties interactively execute $\prod_{\rm rec} (\mathcal{P},\langle e \rangle )$ (Protocol~\ref{pro:rec}) and $\prod_{\rm rec} (\mathcal{P},\langle f \rangle )$ (Protocol~\ref{pro:rec}).
    
        \STATE $P_j$ for $j \in \{0,1\}$ locally computes $\langle z \rangle_j = \langle x \rangle_j \cdot f - \langle v \rangle_j \cdot e+\langle h \rangle_j$.
        Besides, $P_0$ computes $\langle z \rangle_3 = \langle x \rangle_3 \cdot f - \langle v \rangle_3 \cdot e+\langle h \rangle_3$. 
    \end{algorithmic}
\end{algorithm}
\vspace{-2mm}

Protocol \ref{pro:mul2} shows 2PC secure multiplication protocol $\prod_{\rm mul2} (\mathcal{P},\langle x \rangle,
\langle y \rangle)$. 
Firstly, $P_0$ locally computes $\langle e \rangle_0 = \langle x \rangle_0 + \langle u \rangle_0$, $\langle e \rangle_3 = \langle x \rangle_3 + \langle u \rangle_3$ and $\langle f \rangle_0 =\langle y \rangle_0 + \langle v \rangle_0$, $\langle f \rangle_3 =\langle y \rangle_3 + \langle v \rangle_3$. $P_1$ also locally computes $\langle e \rangle_1 = \langle x \rangle_1 + \langle u \rangle_1$ and $\langle f \rangle_1 =\langle y \rangle_1 + \langle v \rangle_1$.
Then $P_0$ and $P_1$ interactively execute $\prod_{\rm rec} (\mathcal{P},\langle e \rangle )$ (Protocol~\ref{pro:rec}) and $\prod_{\rm rec} (\mathcal{P},\langle f \rangle )$ (Protocol~\ref{pro:rec}) to obtain $e$ and $f$ respectively. Finally, $P_0$ computes $\langle z \rangle_0 = \langle x \rangle_0 \cdot f - \langle v \rangle_0 \cdot e+\langle h \rangle_0$, $\langle z \rangle_3 = \langle x \rangle_3 \cdot f - \langle v \rangle_3 \cdot e+\langle h \rangle_3$, and $P_1$ computes $\langle z \rangle_1 = \langle x \rangle_1 \cdot f - \langle v \rangle_1 \cdot e+\langle h \rangle_1$.

\noindent\textbf{Sharing Conversion:}
If $P_2$ drops out, it is trivial to see that the conversions between $\langle \cdot \rangle$-sharing and $[ \cdot ]$-sharing and conversions between $[ \cdot ]$-sharing and $\langle \cdot \rangle$-sharing can be done by $P_0$ and $P_1$ locally.

\begin{itemize}[leftmargin=*]
    \item \emph{Converting $\langle \cdot \rangle${-sharing} to $[ \cdot ]${-sharing}:}
    $P_0$ locally computes $[x]_0=c'_0 \cdot \langle x \rangle_0$ and $[x]_3=c'_3 \cdot \langle x \rangle_3$. Besides, $P_1$ locally computes $[x]_1=c'_1 \cdot \langle x \rangle_1$, such that $x = c'_0\cdot \langle x \rangle_0 + c'_1 \cdot \langle x \rangle_1 + c'_3 \cdot \langle x \rangle_3 = [x]_0+[x]_1+[x]_3$. Therefore, $P_0$ and $P_1$ convert their $\langle \cdot \rangle$-shares to $[ \cdot ]$-shares.

    \item \emph{Converting $[ \cdot ]${-sharing} to $\langle \cdot \rangle${-sharing}:} 
     $P_0$ locally computes $\langle x \rangle_0 = [x]_0 /c'_0$ and $\langle x \rangle_3 = [x]_3 /c'_3$. Besides, $P_1$ locally computes $\langle x \rangle_1 = [x]_1 /c'_1$.
\end{itemize}

\vspace{-2mm}
\begin{algorithm}
    \LinesNumbered
    \small
    \caption{$\prod_{\rm trunc2} (\mathcal{P}, \langle z \rangle)$}
    \label{pro:trunc2}
    \begin{flushleft}
    \textbf{Preprocessing:} Parties pre-shared random values $\langle r \rangle$ and $\langle r' \rangle = \langle r / 2^{\ell_f} \rangle$ \\ 
    \textbf{Input:}  $\langle z \rangle$\\
    \textbf{Output:}  The result after truncation $\langle z' \rangle$, where $z' = z/ 2^{\ell_f}$
    \end{flushleft}
    \begin{algorithmic}[1]
        \STATE $P_j$ for $j \in \{0,1\}$ locally computes $\langle z - r \rangle_j = \langle z \rangle_j - \langle r \rangle_j$. $P_0$ also computes $\langle z - r \rangle_3 = \langle z \rangle_3 - \langle r \rangle_3$;
    
        \STATE $P_1$ sends $\langle z - r \rangle_1$ to $P_0$.
    
        \STATE $P_0$ locally computes $ \langle z' \rangle_0 =(z-r)/(2^{\ell_f} \cdot c'_0) + \langle r' \rangle_0$ and holds $ \langle z' \rangle_3 =\langle r' \rangle_3$. $P_1$ holds $ \langle z' \rangle_1 =\langle r' \rangle_1$.
    \end{algorithmic}
\end{algorithm}
\vspace{-2mm}

\noindent\textbf{Truncation:}
If $P_2$ drops out, Equation (\ref{eq.tru}) can be rewritten as:

\vspace{-2mm}
\begin{equation}
\begin{aligned}
\label{eq.tru2}
    z' &= c'_0 \cdot \frac{(z-r)/c'_0+\langle r \rangle_0}{2^{\ell_f}}+c'_1 \cdot \frac{\langle r \rangle_1}{2^{\ell_f}} +c'_3 \cdot \frac{\langle r \rangle_3}{2^{\ell_f}}
\end{aligned}
\end{equation}

Protocol \ref{pro:trunc2} shows the 2PC secure truncation protocol $\prod_{\rm trunc2} (\mathcal{P}, \langle z \rangle)$. Firstly, $P_1$ sends $\langle z-r \rangle_1$ to $P_0$. Then $P_0$ locally computes $ z-r = c'_0 \cdot \langle z-r \rangle_0 + c'_1 \cdot \langle z - r \rangle_1 + c'_3 \cdot \langle z-r \rangle_3$ and  $(z-r)/(2^{\ell_f} \cdot c'_0) + \langle r' \rangle_0$. Besides, $P_0$ also holds $\langle r' \rangle_3$ and $P_1$ holds $\langle r' \rangle_1$. Note that matrix addition and matrix multiplication protocols for 2PC generalize secure addition and secure multiplication protocols for 2PC. These protocols are similar to the ones for 3PC. In addition, MSB extraction and Bit2A protocols for 2PC are the same as the ones for 3PC.

\begin{table}[ht]
\centering
\caption{Communication rounds and total communication size (bit) cost of building blocks in \pmpl, \texttt{SecureML} and \texttt{TF-Encrypted}. Here, $\ell$ denotes the number of bits of a value. $n\times d, d \times m$ are the sizes for the left and right inputs of matrix-based computations.
$\texttt{ReLU}$ and $\texttt{Sigmoid}$ are executed on a single value. 
$\lambda$ is the security parameter of oblivious transfer used in \texttt{SecureML}.
Rounds stands for online communication rounds and Comm. stands for online communication size.}
\label{table.round}
\renewcommand\arraystretch{1}
\vspace{-2mm}
\scalebox{0.65}{
\begin{tabular}{cccccc}
\toprule
\multirow{2}{*}{\begin{tabular}[c]{@{}c@{}}Building\\ block\end{tabular}}                  & \multirow{2}{*}{Framework} & \multicolumn{2}{c}{3PC}                                          & \multicolumn{2}{c}{2PC}                                                \\ \cline{3-6} 
                                                                                           &                            & \multicolumn{1}{c}{Rounds}        & Comm.                        & \multicolumn{1}{c}{Rounds}        & Comm.                              \\ \hline
\multirow{3}{*}{\begin{tabular}[c]{@{}c@{}}Matrix \\ addition\end{tabular}}                & \pmpl                      & \multicolumn{1}{c}{0}             & 0                            & \multicolumn{1}{c}{0}             & 0                                  \\ 
                                                                                           & \texttt{SecureML}          & \multicolumn{1}{c}{$\backslash$}  & $\backslash$                 & \multicolumn{1}{c}{0}             & 0                                  \\ 
                                                                                           & \texttt{TF-Encrypted}      & \multicolumn{1}{c}{0}             & 0                            & \multicolumn{1}{c}{$\backslash$}  & $\backslash$                       \\ \hline
\multirow{3}{*}{\begin{tabular}[c]{@{}c@{}}Matrix \\ multiplication\end{tabular}}          & \pmpl                      & \multicolumn{1}{c}{1}             & 6$\ell$($nd+dm$)             & \multicolumn{1}{c}{1}             & 3$\ell$($nd+dm$)                   \\ 
                                                                                           & \texttt{SecureML}          & \multicolumn{1}{c}{$\backslash$}  & $\backslash$                 & \multicolumn{1}{c}{1}             & 2$\ell$($nd+dm$)                   \\ 
                                                                                           & \texttt{TF-Encrypted}      & \multicolumn{1}{c}{1}             & 3$\ell$$nm$                  & \multicolumn{1}{c}{$\backslash$}  & $\backslash$                       \\ \hline
\multirow{3}{*}{\begin{tabular}[c]{@{}c@{}}Matrix \\ truncation\end{tabular}}              & \pmpl                      & \multicolumn{1}{c}{1}             & 2$\ell$$nm$                  & \multicolumn{1}{c}{1}             & $\ell$$nm$                         \\ 
                                                                                           & \texttt{SecureML}          & \multicolumn{1}{c}{$\backslash$}  & $\backslash$                 & \multicolumn{1}{c}{0}             & 0                                  \\ 
                                                                                           & \texttt{TF-Encrypted}      & \multicolumn{1}{c}{1}             & 2$\ell$$nm$                  & \multicolumn{1}{c}{$\backslash$}  & $\backslash$                       \\ \hline
\multirow{3}{*}{\begin{tabular}[c]{@{}c@{}}Multiplication \\ with truncation\end{tabular}} & \pmpl                      & \multicolumn{1}{c}{2}             & 6$\ell$($nd+dm$)+2$\ell$$nm$ & \multicolumn{1}{c}{2}             & $\ell$$nm$+3$\ell$($nd+dm$)        \\ 
                                                                                           & \texttt{SecureML}          & \multicolumn{1}{c}{$\backslash$}  & $\backslash$                 & \multicolumn{1}{c}{1}             & 2$\ell$($nd+dm$)                   \\ 
                                                                                           & \texttt{TF-Encrypted}      & \multicolumn{1}{c}{1}             & 4$\ell$$nm$                  & \multicolumn{1}{c}{$\backslash$}  & $\backslash$                       \\ \hline
\multirow{3}{*}{\texttt{ReLU}}                                                             & \pmpl                      & \multicolumn{1}{c}{$\log\ell$+5}  & $18 \ell+4\ell \log \ell$    & \multicolumn{1}{c}{$\log\ell $+4} & $8 \ell +2 \ell \log\ell$          \\ 
                                                                                           & \texttt{SecureML}          & \multicolumn{1}{c}{$\backslash$}  & $\backslash$                 & \multicolumn{1}{c}{2}             & $4\lambda(\ell-1)+2(\ell+\lambda)$ \\  
                                                                                           & \texttt{TF-Encrypted}      & \multicolumn{1}{c}{$\log\ell$+1}  & $3 \ell+3\ell \log\ell$      & \multicolumn{1}{c}{$\backslash$}  & $\backslash$                       \\ \hline
\multirow{3}{*}{\texttt{Sigmoid}}                                                          & \pmpl                      & \multicolumn{1}{c}{$\log \ell$+6} & $38 \ell +8 \ell \log \ell$  & \multicolumn{1}{c}{$\log \ell$+5} & $18 \ell+4 \ell \log\ell$          \\ 
                                                                                           & \texttt{SecureML}          & \multicolumn{1}{c}{$\backslash$}  & $\backslash$                 & \multicolumn{1}{c}{4}             & $4\lambda(2\ell-1)+6\ell$          \\ 
                                                                                           & \texttt{TF-Encrypted}      & \multicolumn{1}{c}{$\log\ell$+3}  & $9 \ell+3\ell \log\ell$      & \multicolumn{1}{c}{$\backslash$}  & $\backslash$                       \\ \bottomrule
\end{tabular}}
\vspace{-3mm}
\end{table}

\vspace{-2mm}
\subsection{Complexity Analysis}
We measure the cost of each building block from two aspects: online communication rounds and online communication size in both 3PC (no party drops out) and 2PC ($P_2$ drops out) settings. Table~\ref{table.round} shows the comparison of the communication rounds and communication size among \pmpl, \texttt{SecureML} and \texttt{TF-Encrypted}.

\vspace{-2mm}
\section{Evaluation}
\label{sec.evaluation}

In this section, we present the implementation of linear regression, logistic regression and neural networks in \pmpl. Meanwhile, we conduct experiments to evaluate the performance of \pmpl by the comparison with other MPL frameworks.

\vspace{-2mm}
\subsection{Experiment Settings and Datasets}
\label{sec.settings}

\noindent\textbf{Experiment Settings:}
We conduct 3PC experiments on three Linux servers equipped with 20-core 2.4 Ghz Intel Xeon CPUs and 128GB of RAM, and 2PC experiments on two Linux servers equipped same as above. 
The experiments are performed on two network environments: one is the LAN setting with a bandwidth of 1Gbps and sub-millisecond RTT (round-trip time) latency, the other one is the WAN setting with 40MBps bandwidth and 40ms RTT latency.
Note that we run \texttt{TF-Encrypted} (with \texttt{ABY3} as the back-end framework) under the above environment. While the experimental results of \texttt{SecureML} are from the study~\cite{mohassel2017secureml} and~\cite{mohassel2018aby3} since the code of
\texttt{SecureML} is not public.
We implement \pmpl in C++ over the ring $\mathbb{Z}_{2^\ell}$.
Here, we set $\ell =$ 64, and the least 20 significant bits $\ell_f$ represent the fractional part, which is the same as the setting of \texttt{SecureML} and \texttt{TF-Encrypted}.
Additionally, we set \textit{public matrix} $\varPhi(\mathcal{P})$ as follows:
\begin{equation*}
    \varPhi (\mathcal{P}) = 
\begin{bmatrix}
    \varPhi(0) \\ 
    \varPhi(1) \\ 
    \varPhi(2) \\ 
    \varPhi(3)
\end{bmatrix}
=
\begin{bmatrix}
    1 & 0  & 1 \\ 
    1 & 1  & 2^\ell-1 \\ 
    2 & 2  & 2^\ell-3 \\ 
    3 & 3  & 2^\ell-4
\end{bmatrix}
\end{equation*}
Therefore, according to Equation (\ref{eq.coefficient}), we can compute $c_0 = 1, c_1 = 2^\ell-2, c_2 = 1, c'_0 = 1, c'_1 = 2^\ell-3, c'_3 = 1, c''_0 = 1, c''_2 = 3, c''_3 = 2^\ell-2$.

\noindent\textbf{Datasets:}
To evaluate the performance of \pmpl, we use the MNIST dataset\cite{DBLP:journals/pieee/LeCunBBH98}.
It contains image samples of handwritten digits from ``0'' to ``9'', each with 784 features representing 28 × 28 pixels. Besides, the greyscale of each pixel is between 0$\sim$255. Its training set contains 60,000 samples, and the testing set contains 10,000 samples. 
For linear regression and logistic regression, we consider binary classification, where the digits "$0$" as a class, and the digits "$1\sim9$" as another one. For BP neural network, we consider a ten-class classification task. Additionally, we benchmark more complex datasets, including Fashion-MNIST~\cite{DBLP:journals/corr/abs-1708-07747}
and SVHN~\cite{netzer2011reading}, in Appendix~\ref{sec.accuracy}.

\subsection{Offline Phase}
We evaluate the performance of generating the vector multiplication triplets under the LAN setting in the offline phase. We follow the same setting as \texttt{SecureML}, where the batch size $B=128$, epoch $E=2$, the number of samples $n \in \{ 100, 1,000, 10,000 \}$ and the dimension $D \in \{100, 500, 1,000\}$. The number of iterations is $n*E /B$.  As is shown in Table \ref{table.offline}, \pmpl is faster than both \texttt{SecureML} based on HE protocol and \texttt{SecureML} based on OT protocol. Especially when the dimension $D = 1,000$ and number of samples $n= 10,000 $, \pmpl is around 119$\times$ faster than \texttt{SecureML} based on HE protocol and around 6$\times$ faster than \texttt{SecureML} based on OT protocol.

\vspace{-2mm}
\begin{table}[h]
\centering
\caption{Performance of the offline phase (\textit{seconds}). $*$ means estimated via extrapolation.}
\label{table.offline}
\renewcommand\arraystretch{1}
\vspace{-2mm}
\scalebox{0.75}{
\begin{tabular}{ccccc}
\toprule
\multirow{2}{*}{Number of samples $n$} & \multirow{2}{*}{Protocol}    & \multicolumn{3}{c}{Dimension ($D$)}                                   \\ \cline{3-5} 
                                      &                              & \multicolumn{1}{c}{100}    & \multicolumn{1}{c}{500}    & 1,000      \\ \hline
\multirow{3}{*}{1,000}                & \pmpl                        & \multicolumn{1}{c}{0.34}   & \multicolumn{1}{c}{0.78}   & 1.33       \\ 
                                      & \texttt{SecureML} (HE-based) & \multicolumn{1}{c}{23.9}   & \multicolumn{1}{c}{83.9}   & 158.4      \\ 
                                      & \texttt{SecureML}(OT-based)  & \multicolumn{1}{c}{0.86}   & \multicolumn{1}{c}{3.8}    & 7.9        \\ \hline
\multirow{3}{*}{10,000}               & \pmpl                        & \multicolumn{1}{c}{3.73}   & \multicolumn{1}{c}{7.89}   & 13.21      \\ 
                                      & \texttt{SecureML} (HE-based) & \multicolumn{1}{c}{248.4}  & \multicolumn{1}{c}{869.1}  & 1600.9     \\ 
                                      & \texttt{SecureML}(OT-based)  & \multicolumn{1}{c}{7.9}    & \multicolumn{1}{c}{39.2}   & 80.0       \\ \hline
\multirow{3}{*}{100,000}              & \pmpl                        & \multicolumn{1}{c}{38.05}  & \multicolumn{1}{c}{78.70}  & 140.28     \\ 
                                      & \texttt{SecureML} (HE-based) & \multicolumn{1}{c}{2437.1} & \multicolumn{1}{c}{8721.5} & 16000$^*$ \\ 
                                      & \texttt{SecureML}(OT-based)  & \multicolumn{1}{c}{88.0}   & \multicolumn{1}{c}{377.9}  & 794.0      \\ \bottomrule
\end{tabular}}
\vspace{-3mm}
\end{table}

\subsection{Secure Training in Online Phase}
\label{sec.online}
As is mentioned in Section \ref{sec.ml},  the training of the evaluated machine learning models consists of two phases: (1) the forward propagation phase is to compute the output; (2) the backward propagation phase is to update coefficient parameters according to the error between the output computed in the forward propagation and the actual label. One iteration in the training phase contains one forward propagation and a backward propagation.

To compare \pmpl with \texttt{SecureML} and \texttt{TF-Encrypted}, we select $D \in \{10, 100, 1,000\}$ and $B \in \{128, 256, 512, 1,024\}$. 
In addition, we consider two scenarios for experiments, i.e. 3PC with no \textit{assistant party} drops out, and 2PC with $P_2$ drops out.

\noindent\textbf{Linear Regression:}
We use mini-batch stochastic gradient descent (SGD for short) to train a linear regression model. The update function in Equation (\ref{eq.linear_w}) can be expressed as:

\begin{equation*}
    \vec{w} :=\vec{w} - \frac{\alpha}{B} \mathbf{X}_i^{T} \times (\mathbf{X}_i \times \vec{w}-\mathbf{\mathbf{Y}_i})
\end{equation*}
where $\mathbf{X_i}$ is a subset of batch size $B$. Besides, $(\mathbf{X}_i, \mathbf{Y}_i)$ are randomly selected from the whole dataset in the $i$-th iteration.

\begin{table}[t]
\centering
\caption{Online throughput of linear regression compared to \texttt{SecureML} and \texttt{TF-Encrypted} (iterations/second). }
\label{table.linear}
\renewcommand\arraystretch{1}
\vspace{-2mm}
\scalebox{0.75}{
\begin{tabular}{ccccccc}
\toprule
\multirow{2}{*}{Setting} & \multirow{2}{*}{\begin{tabular}[c]{@{}c@{}}Dimension \\ ($D$)\end{tabular}} & \multirow{2}{*}{Protocol} & \multicolumn{4}{c}{Batch Size ($B$)}                                                                \\ \cline{4-7} 
                         &                                                                             &                           & \multicolumn{1}{c}{128}     & \multicolumn{1}{c}{256}     & \multicolumn{1}{c}{512}     & 1,024   \\ \hline
\multirow{12}{*}{LAN}    & \multirow{4}{*}{10}                                                         & \pmpl (3PC)               & \multicolumn{1}{c}{4545.45} & \multicolumn{1}{c}{3846.15} & \multicolumn{1}{c}{2631.58} & 1666.67 \\  
                         &                                                                             & \pmpl (2PC)               & \multicolumn{1}{c}{5263.16} & \multicolumn{1}{c}{4166.67} & \multicolumn{1}{c}{2777.78} & 1694.92 \\  
                         &                                                                             & \texttt{SecureML}         & \multicolumn{1}{c}{7,889}   & \multicolumn{1}{c}{7,206}   & \multicolumn{1}{c}{4,350}   & 4,263   \\  
                         &                                                                             & \texttt{TF-Encrypted}     & \multicolumn{1}{c}{282.36}  & \multicolumn{1}{c}{248.47}  & \multicolumn{1}{c}{195.18}  & 139.51  \\ \cline{2-7} 
                         & \multirow{4}{*}{100}                                                        & \pmpl (3PC)               & \multicolumn{1}{c}{1333.33} & \multicolumn{1}{c}{740.74}  & \multicolumn{1}{c}{387.60}  & 166.67  \\  
                         &                                                                             & \pmpl (2PC)               & \multicolumn{1}{c}{1428.57} & \multicolumn{1}{c}{813.01}  & \multicolumn{1}{c}{436.68}  & 202.02  \\  
                         &                                                                             & \texttt{SecureML}         & \multicolumn{1}{c}{2,612}   & \multicolumn{1}{c}{755}     & \multicolumn{1}{c}{325}     & 281     \\  
                         &                                                                             & \texttt{TF-Encrypted}     & \multicolumn{1}{c}{141.17}  & \multicolumn{1}{c}{90.95}   & \multicolumn{1}{c}{55.36}   & 30.06   \\ \cline{2-7} 
                         & \multirow{4}{*}{1,000}                                                      & \pmpl (3PC)               & \multicolumn{1}{c}{89.05}   & \multicolumn{1}{c}{39.53}   & \multicolumn{1}{c}{17.74}   & 8.87    \\  
                         &                                                                             & \pmpl (2PC)               & \multicolumn{1}{c}{137.36}  & \multicolumn{1}{c}{58.82}   & \multicolumn{1}{c}{26.39}   & 12.43   \\  
                         &                                                                             & \texttt{SecureML}         & \multicolumn{1}{c}{131}     & \multicolumn{1}{c}{96}      & \multicolumn{1}{c}{45}      & 27      \\  
                         &                                                                             & \texttt{TF-Encrypted}     & \multicolumn{1}{c}{24.53}   & \multicolumn{1}{c}{12.74}   & \multicolumn{1}{c}{6.55}    & 3.30    \\ \hline
\multirow{12}{*}{WAN}    & \multirow{4}{*}{10}                                                         & \pmpl (3PC)               & \multicolumn{1}{c}{4.93}    & \multicolumn{1}{c}{4.89}    & \multicolumn{1}{c}{4.84}    & 4.73    \\  
                         &                                                                             & \pmpl (2PC)               & \multicolumn{1}{c}{4.94}    & \multicolumn{1}{c}{4.921}   & \multicolumn{1}{c}{4.88}    & 4.80    \\  
                         &                                                                             & \texttt{SecureML}         & \multicolumn{1}{c}{12.40}   & \multicolumn{1}{c}{12.40}   & \multicolumn{1}{c}{12.40}   & 12.40   \\  
                         &                                                                             & \texttt{TF-Encrypted}     & \multicolumn{1}{c}{11.58}   & \multicolumn{1}{c}{11.53}   & \multicolumn{1}{c}{11.42}   & 11.15   \\ \cline{2-7} 
                         & \multirow{4}{*}{100}                                                        & \pmpl (3PC)               & \multicolumn{1}{c}{4.66}    & \multicolumn{1}{c}{4.47}    & \multicolumn{1}{c}{4.10}    & 3.55    \\  
                         &                                                                             & \pmpl (2PC)               & \multicolumn{1}{c}{4.75}    & \multicolumn{1}{c}{4.67}    & \multicolumn{1}{c}{4.30}    & 4.03    \\  
                         &                                                                             & \texttt{SecureML}         & \multicolumn{1}{c}{12.30}   & \multicolumn{1}{c}{12.20}   & \multicolumn{1}{c}{11.80}   & 11.80   \\  
                         &                                                                             & \texttt{TF-Encrypted}     & \multicolumn{1}{c}{11.13}   & \multicolumn{1}{c}{10.63}   & \multicolumn{1}{c}{9.74}    & 8.32    \\ \cline{2-7} 
                         & \multirow{4}{*}{1,000}                                                       & \pmpl (3PC)               & \multicolumn{1}{c}{3.29}    & \multicolumn{1}{c}{2.47}    & \multicolumn{1}{c}{1.51}    & 0.84    \\  
                         &                                                                             & \pmpl (2PC)               & \multicolumn{1}{c}{3.83}    & \multicolumn{1}{c}{3.14}    & \multicolumn{1}{c}{2.11}    & 1.32    \\  
                         &                                                                             & \texttt{SecureML}         & \multicolumn{1}{c}{11.00}   & \multicolumn{1}{c}{9.80}    & \multicolumn{1}{c}{9.20}    & 7.30    \\  
                         &                                                                             & \texttt{TF-Encrypted}     & \multicolumn{1}{c}{7.85}    & \multicolumn{1}{c}{5.76}    & \multicolumn{1}{c}{3.80}    & 2.22    \\ \bottomrule
\end{tabular}
}
\vspace{-3mm}
\end{table}

As is shown in Table \ref{table.linear}, the experimental results show that:

(1) In the LAN setting, \pmpl for 3PC is around 2.7$\times$ $\sim$ 16.1$\times$ faster and \pmpl for 2PC is around 3.8$\times \sim$ 18.6$\times$ faster than \texttt{TF-Encrypted}. We analyze that this is due to \texttt{Tensorflow}, which is the basis of \texttt{TF-Encrypted}, bringing some extra overhead, e.g. operator schedulings. As the training process of linear regression is relatively simple, when we train linear regression with \texttt{TF-Encrypted}, the extra overhead brought by \texttt{Tensorflow} becomes the main performance bottleneck.
Besides, \texttt{SecureML} is faster than \pmpl. The performance differences between \pmpl and \texttt{SecureML} are led by two reasons.  First of all, the experiment environments are different. As the source code of \texttt{SecureML} is not available, the experimental results of \texttt{SecureML}, which are obtained in the different environment with \pmpl, are from the study~\cite{mohassel2018aby3}. More specifically, we perform our experiment on 2.4 Ghz Intel Xeon CPUs and 128GB of RAM, while the study~\cite{mohassel2018aby3} performs on 2.7 Ghz Intel Xeon CPUs and 256GB of RAM, which leads to the local computing of \texttt{SecureML} being faster than \pmpl. Meanwhile, our bandwidth is 1Gbps, while the bandwidth of the study~\cite{mohassel2018aby3} is 10 Gbps. 
Second, the underlying techniques are different. The online communication overhead of building blocks in \pmpl is more than those in \texttt{SecureML} (as shown in Table~\ref{table.round}). For instance, the truncation operation in \pmpl needs one round while \texttt{SecureML} performs the truncation operation locally without communication. 

(2) In the WAN setting, \texttt{SecureML} and \texttt{TF-Encrypted} are faster than \pmpl. This is because to provide more security guarantees (i.e., defending the collusion of two assistant parties) and ensure robustness, \pmpl requires more communication overhead than \texttt{SecureML} and \texttt{TF-Encrypted} (as shown in Table~\ref{table.round}). Therefore, the performance of \pmpl is promising.

(3) In the both LAN setting and WAN setting, \pmpl for 2PC is faster than 3PC. This is because the communication overhead of 2PC is smaller. 

Besides, the trained model can reach an accuracy of 97\% on the test dataset.

\begin{table}[htb]
\centering
\caption{Online throughput of logistic regression compared to \texttt{SecureML} and \texttt{TF-Encrypted} (iterations/second).} 
\vspace{-2mm}
\label{table.logistic}
\renewcommand\arraystretch{1}
\scalebox{0.75}{
\begin{tabular}{ccccccc}
\toprule
\multirow{2}{*}{Setting} & \multirow{2}{*}{\begin{tabular}[c]{@{}c@{}}Dimension \\ ($D$)\end{tabular}} & \multirow{2}{*}{Protocol} & \multicolumn{4}{c}{Batch Size ($B$)}                                                            \\ \cline{4-7} 
                         &                                                                             &                           & \multicolumn{1}{c}{128}    & \multicolumn{1}{c}{256}    & \multicolumn{1}{c}{512}    & 1,024  \\ \hline
\multirow{12}{*}{LAN}    & \multirow{4}{*}{10}                                                         & \pmpl (3PC)               & \multicolumn{1}{c}{579.45} & \multicolumn{1}{c}{537.47} & \multicolumn{1}{c}{444.45} & 330.40 \\ 
                         &                                                                             & \pmpl (2PC)               & \multicolumn{1}{c}{598.75} & \multicolumn{1}{c}{542.68} & \multicolumn{1}{c}{455.19} & 332.68 \\ \
                         &                                                                             & \texttt{SecureML}         & \multicolumn{1}{c}{188}    & \multicolumn{1}{c}{101}    & \multicolumn{1}{c}{41}     & 25     \\  
                         &                                                                             & \texttt{TF-Encrypted}     & \multicolumn{1}{c}{119.88} & \multicolumn{1}{c}{110.78} & \multicolumn{1}{c}{97.16}  & 74.07  \\ \cline{2-7} 
                         & \multirow{4}{*}{100}                                                        & \pmpl (3PC)               & \multicolumn{1}{c}{425.88} & \multicolumn{1}{c}{332.86} & \multicolumn{1}{c}{222.89} & 121.92 \\ 
                         &                                                                             & \pmpl (2PC)               & \multicolumn{1}{c}{435.41} & \multicolumn{1}{c}{353.55} & \multicolumn{1}{c}{235.93} & 128.25 \\ 
                         &                                                                             & \texttt{SecureML}         & \multicolumn{1}{c}{183}    & \multicolumn{1}{c}{93}     & \multicolumn{1}{c}{46}     & 24     \\  
                         &                                                                             & \texttt{TF-Encrypted}     & \multicolumn{1}{c}{87.34}  & \multicolumn{1}{c}{63.06}  & \multicolumn{1}{c}{41.25}  & 25.12  \\ \cline{2-7} 
                         & \multirow{4}{*}{1,000}                                                      & \pmpl (3PC)               & \multicolumn{1}{c}{100.66} & \multicolumn{1}{c}{49.53}  & \multicolumn{1}{c}{22.85}  & 11.18  \\  
                         &                                                                             & \pmpl (2PC)               & \multicolumn{1}{c}{105.82} & \multicolumn{1}{c}{51.62}  & \multicolumn{1}{c}{23.37}  & 11.40  \\ 
                         &                                                                             & \texttt{SecureML}         & \multicolumn{1}{c}{105}    & \multicolumn{1}{c}{51}     & \multicolumn{1}{c}{24}     & 13.50  \\ 
                         &                                                                             & \texttt{TF-Encrypted}     & \multicolumn{1}{c}{22.10}  & \multicolumn{1}{c}{12.07}  & \multicolumn{1}{c}{6.42}   & 3.28   \\ \hline
\multirow{12}{*}{WAN}    & \multirow{4}{*}{10}                                                         & \pmpl (3PC)               & \multicolumn{1}{c}{0.65}   & \multicolumn{1}{c}{0.64}   & \multicolumn{1}{c}{0.63}   & 0.62   \\ 
                         &                                                                             & \pmpl (2PC)               & \multicolumn{1}{c}{0.65}   & \multicolumn{1}{c}{0.65}   & \multicolumn{1}{c}{0.64}   & 0.63   \\ 
                         &                                                                             & \texttt{SecureML}         & \multicolumn{1}{c}{3.10}   & \multicolumn{1}{c}{2.28}   & \multicolumn{1}{c}{1.58}   & 0.99   \\
                         &                                                                             & \texttt{TF-Encrypted}     & \multicolumn{1}{c}{4.92}   & \multicolumn{1}{c}{4.91}   & \multicolumn{1}{c}{4.90}   & 4.81   \\ \cline{2-7} 
                         & \multirow{4}{*}{100}                                                        & \pmpl (3PC)               & \multicolumn{1}{c}{0.63}   & \multicolumn{1}{c}{0.62}   & \multicolumn{1}{c}{0.60}   & 0.56   \\ 
                         &                                                                             & \pmpl (2PC)               & \multicolumn{1}{c}{0.64}   & \multicolumn{1}{c}{0.63}   & \multicolumn{1}{c}{0.62}   & 0.60   \\ 
                         &                                                                             & \texttt{SecureML}         & \multicolumn{1}{c}{3.08}   & \multicolumn{1}{c}{2.25}   & \multicolumn{1}{c}{1.57}   & 0.99   \\ 
                         &                                                                             & \texttt{TF-Encrypted}     & \multicolumn{1}{c}{4.83}   & \multicolumn{1}{c}{4.69}   & \multicolumn{1}{c}{4.59}   & 4.21   \\ \cline{2-7} 
                         & \multirow{4}{*}{1,000}                                                      & \pmpl (3PC)               & \multicolumn{1}{c}{0.56}   & \multicolumn{1}{c}{0.52}   & \multicolumn{1}{c}{0.42}   & 0.32   \\ 
                         &                                                                             & \pmpl (2PC)               & \multicolumn{1}{c}{0.60}   & \multicolumn{1}{c}{0.57}   & \multicolumn{1}{c}{0.51}   & 0.42   \\ 
                         &                                                                             & \texttt{SecureML}         & \multicolumn{1}{c}{3.01}   & \multicolumn{1}{c}{2.15}   & \multicolumn{1}{c}{1.47}   & 0.93   \\ 
                         &                                                                             & \texttt{TF-Encrypted}     & \multicolumn{1}{c}{4.05}   & \multicolumn{1}{c}{3.47}   & \multicolumn{1}{c}{2.65}   & 1.76   \\ \bottomrule
\end{tabular}}
\vspace{-3mm}
\end{table}

\noindent\textbf{Logistic Regression:}
Similar to linear regression, the update function using mini-batch SGD method in logistic regression can be expressed as:
\begin{equation*}
    \vec{w} :=\vec{w} - \frac{\alpha}{B} \mathbf{X}_i^{T} \times (\texttt{Sigmoid} (\mathbf{X}_i \times \vec{w})-\mathbf{\mathbf{Y}_i})
\end{equation*}

As is shown in Table \ref{table.logistic}, the experimental results show that: 

(1) In the LAN setting, \pmpl is faster than both \texttt{SecureML} and  \texttt{TF-Encrypted}. 
The reason for these performance differences between \pmpl and \texttt{SecureML} is \texttt{SecureML} implements \texttt{Sigmoid} utilizing the garbled circuit and oblivious transfer.  It requires fewer communication rounds but much bigger communication size than those in \pmpl (as shown in Table~\ref{table.round}).
Besides, the reasons for these performance differences between \pmpl and \texttt{TF-Encrypted} are the same as those for linear regression. 

(2) In the WAN setting, \texttt{SecureML} and \texttt{TF-Encrypted} are faster than \pmpl. This is because the communication rounds are important performance bottlenecks in the WAN setting. Meanwhile, \pmpl requires more communication rounds than \texttt{SecureML} and \texttt{TF-Encrypted} (as shown in Table~\ref{table.round}) to provide more security guarantees (i.e., defending the collusion of two assist parties) and ensure robustness. Therefore, the performance of \pmpl is promising.

(3) \pmpl for 2PC is faster than 3PC. This is also because the communication overhead of 2PC is smaller.

Besides, the trained model can reach an accuracy of 99\% on the test dataset.

\noindent\textbf{BP Neural Networks:}
For BP neural networks, we follow the steps similar to those of \texttt{SecureML} and \texttt{TF-Encrypted}. 
In \pmpl, we consider a classical BP neural network consisting of four layers, including one input layer, two hidden layers, and one output layer. Besides, we use \texttt{ReLU} as the activation function. 
As is shown in Table \ref{table.nn}, the experimental results show that: 

(1) \texttt{TF-Encrypted} is faster than \pmpl. 
When we train BP neural networks, which are more complex than linear regression and logistic regression, the overhead of model training becomes the performance bottleneck in \texttt{TF-Encrypted} rather than the extra overhead brought by \texttt{Tensorflow}.
Meanwhile, \pmpl requires more communication overhead (as shown in Table~\ref{table.round}) than \texttt{TF-Encrypted} to provide more security guarantees (i.e., defending the collusion of two assist parties) and ensure robustness, two requirements from novel practical scenarios. The performance of \pmpl is still promising.

(2) \pmpl for 2PC is faster than 3PC. This is also because the communication overhead of 2PC is smaller.

After training the neural network on MNIST dataset with batch size $B=128$, dimension $D=784$, \pmpl can reach the accuracy of 96\% on the test dataset.

\begin{table}[ht]
\centering
\caption{Online throughput of BP neural networks compared to \texttt{TF-Encrypted} (iterations/second).} 
\label{table.nn}
\renewcommand\arraystretch{1}
\vspace{-2mm}
\scalebox{0.8}{
\begin{tabular}{ccccccc}
\toprule
\multirow{2}{*}{Setting} & \multirow{2}{*}{\begin{tabular}[c]{@{}c@{}}Dimension \\ ($D$)\end{tabular}} & \multirow{2}{*}{Protocol} & \multicolumn{4}{c}{Batch Size ($B$)}                                                        \\ \cline{4-7} 
                         &                                                                             &                           & \multicolumn{1}{c}{128}   & \multicolumn{1}{c}{256}   & \multicolumn{1}{c}{512}   & 1,024 \\ \hline
\multirow{9}{*}{LAN}     & \multirow{3}{*}{10}                                                         & \pmpl (3PC)               & \multicolumn{1}{c}{16.49} & \multicolumn{1}{c}{8.43}  & \multicolumn{1}{c}{4.08}  & 1.86  \\  
                         &                                                                             & \pmpl (2PC)               & \multicolumn{1}{c}{17.61} & \multicolumn{1}{c}{8.62}  & \multicolumn{1}{c}{4.14}  & 1.91  \\
                         &                                                                             & \texttt{TF-Encrypted}     & \multicolumn{1}{c}{29.56} & \multicolumn{1}{c}{18.95} & \multicolumn{1}{c}{11.38} & 6.13  \\ \cline{2-7} 
                         & \multirow{3}{*}{100}                                                        & \pmpl (3PC)               & \multicolumn{1}{c}{15.79} & \multicolumn{1}{c}{7.88}  & \multicolumn{1}{c}{3.84}  & 1.77  \\ 
                         &                                                                             & \pmpl (2PC)               & \multicolumn{1}{c}{16.23} & \multicolumn{1}{c}{8.17}  & \multicolumn{1}{c}{3.95}  & 1.81  \\ 
                         &                                                                             & \texttt{TF-Encrypted}     & \multicolumn{1}{c}{25.39} & \multicolumn{1}{c}{15.78} & \multicolumn{1}{c}{8.63}  & 5.02  \\ \cline{2-7} 
                         & \multirow{3}{*}{1,000}                                                      & \pmpl (3PC)               & \multicolumn{1}{c}{8.93}  & \multicolumn{1}{c}{5.25}  & \multicolumn{1}{c}{2.65}  & 1.29  \\ 
                         &                                                                             & \pmpl (2PC)               & \multicolumn{1}{c}{9.19}  & \multicolumn{1}{c}{5.33}  & \multicolumn{1}{c}{2.66}  & 1.31  \\ 
                         &                                                                             & \texttt{TF-Encrypted}     & \multicolumn{1}{c}{12.38} & \multicolumn{1}{c}{6.89}  & \multicolumn{1}{c}{3.54}  & 1.80  \\ \hline
\multirow{9}{*}{WAN}     & \multirow{3}{*}{10}                                                         & \pmpl (3PC)               & \multicolumn{1}{c}{0.15}  & \multicolumn{1}{c}{0.12}  & \multicolumn{1}{c}{0.10}  & 0.07  \\ 
                         &                                                                             & \pmpl (2PC)               & \multicolumn{1}{c}{0.16}  & \multicolumn{1}{c}{0.14}  & \multicolumn{1}{c}{0.12}  & 0.09  \\ 
                         &                                                                             & \texttt{TF-Encrypted}     & \multicolumn{1}{c}{0.93}  & \multicolumn{1}{c}{0.65}  & \multicolumn{1}{c}{0.40}  & 0.22  \\ \cline{2-7} 
                         & \multirow{3}{*}{100}                                                        & \pmpl (3PC)               & \multicolumn{1}{c}{0.15}  & \multicolumn{1}{c}{0.12}  & \multicolumn{1}{c}{0.10}  & 0.07  \\  
                         &                                                                             & \pmpl (2PC)               & \multicolumn{1}{c}{0.16}  & \multicolumn{1}{c}{0.14}  & \multicolumn{1}{c}{0.12}  & 0.09  \\ 
                         &                                                                             & \texttt{TF-Encrypted}     & \multicolumn{1}{c}{0.92}  & \multicolumn{1}{c}{0.64}  & \multicolumn{1}{c}{0.39}  & 0.21  \\ \cline{2-7} 
                         & \multirow{3}{*}{1,000}                                                      & \pmpl (3PC)               & \multicolumn{1}{c}{0.14}  & \multicolumn{1}{c}{0.12}  & \multicolumn{1}{c}{0.09}  & 0.06  \\ 
                         &                                                                             & \pmpl (2PC)               & \multicolumn{1}{c}{0.15}  & \multicolumn{1}{c}{0.13}  & \multicolumn{1}{c}{0.11}  & 0.08  \\ 
                         &                                                                             & \texttt{TF-Encrypted}     & \multicolumn{1}{c}{0.80}  & \multicolumn{1}{c}{0.55}  & \multicolumn{1}{c}{0.33}  & 0.18  \\ \bottomrule
\end{tabular} }
\vspace{-3mm}
\end{table}

\vspace{-2mm}
\section{Discussion}
\label{sec.discussion}

\subsection{\pmpl with More Assistant Parties}
\label{sec.moreparty}
Our proposed \pmpl can be extended to support more \textit{assistant parties} by setting \textit{pubic matrix} $\varPhi (\mathcal{P})$.
In order to support more \textit{assistant parties}, we can increase the number of columns of the \textit{public matrix} $\varPhi (\mathcal{P})$, i.e. expand the dimension of each \textit{public vector} ${\varPhi(i)}$.
For instance, for a set of parties $\mathcal{P} = \{P_0, P_1, P_2, P_3, P_4 \}$ and an access structure $\varGamma = \{ B_0, B_1,  B_2, B_3, B_4 \}$ $= \{ \{P_0, P_1, P_2, P_3, P_4 \},$ $\{P_0, P_2, P_3, P_4\},$ $\{ P_0, P_1, P_3, P_4 \},$ $\{ P_0, P_1, P_2, P_4 \},$ $\{ P_0, P_1, P_2, P_3 \} \}$, where $P_0$ is the \textit{privileged party} and $P_1,P_2,P_3,P_4$ are \textit{assistant parties}. The secret cannot be revealed without the participation of the \textit{privileged party} $P_0$, even when \textit{assistant parties} collude and one of \textit{assistant parties} drops out during training.

To securely perform the training in the above application scenario, the \textit{public matrix} $\varPhi (\mathcal{P})$ with the size of $6 \times 5$ should satisfy the following four restrictions:
\begin{itemize}[leftmargin=*]
    \item $(1,0,0,0,0)$ can be written as a linear combination of \textit{public vectors} in the set $ { \{\varPhi(0)}, {\varPhi(1)}, {\varPhi(2)}, {\varPhi(3)}, {\varPhi(4)} \}$ , where all \textit{public vectors} are linear independent.
    
    \item The alternate \textit{public vector} ${\varPhi(5)}$ held by the \textit{privileged party} $P_0$ can be represented linearly by \textit{public vectors} ${\varPhi(1)}, {\varPhi(2)}, {\varPhi(3)}$ and ${\varPhi(4)}$. That is, ${\varPhi(5)} = \sum \nolimits_{j=1}^4 a_j *{\varPhi(j)}$, where $j \in \{1,2,3,4\}$ and $a_j \neq 0$. Therefore, $(1,0,0,0,0)$ can also be a linear combination of the \textit{public vectors} in sets $ \{ \varPhi(0), \varPhi(2), \varPhi(3), \varPhi(4), \varPhi(5)\}$, $ \{ \varPhi(0), \varPhi(1), \varPhi(3), \varPhi(4), \varPhi(5)\}$, $ \{ \varPhi(0), \varPhi(1), \varPhi(2), \varPhi(4), \varPhi(5)\}$, $ \{ \varPhi(0), \varPhi(1), \varPhi(2), \varPhi(3), \varPhi(5)\}$, respectively. 
    
    \item To guarantee that only the set of parties in the access structure can collaboratively reveal the secret value, $(1,0,0,0,0)$ cannot be represented as a linear combination of \textit{public vectors} in the sets $\{ {\varPhi(1)}, {\varPhi(2)}, {\varPhi(3)},{\varPhi(4)},{\varPhi(5)}\}$, $\{ {\varPhi(0)}, {\varPhi(5)}\}$ and their subsets.
    
    \item The values of \textit{public matrix} $\varPhi (\mathcal{P})$ and reconstruction coefficients should be elements of the ring $\mathbb{Z}_{2^\ell}$.
    
\end{itemize}

For example, a \textit{public matrix} $\varPhi (\mathcal{P})$ that satisfies the above restrictions is:

\begin{equation*}
    \varPhi (\mathcal{P}) = 
\begin{bmatrix}
    {\varPhi(0)} \\ 
    {\varPhi(1)} \\ 
    {\varPhi(2)} \\ 
    {\varPhi(3)} \\
    {\varPhi(4)} \\
    {\varPhi(5)}
\end{bmatrix}
=
\begin{bmatrix}
    1 & 2  & 1 & 2 & 1 \\ 
    2^\ell-1 & 1  & 0 & 1 & 3 \\ 
    1 & 1  & 1 & 0 & 1 \\ 
    0 & 0  & 0 & 2 & 3 \\
    0 & 0  & 0 & 1 & 2 \\
    0 & 2  & 1 & 4 & 9 \\
\end{bmatrix}.
\end{equation*}

Note that we can hereby tolerate more \textit{assistant parties} ($\leq 3$) dropping out during the training by setting more alternate vectors for the \textit{privileged party} $P_0$.
Furthermore, when more \textit{assistant parties} are involved, the protocols proposed in Section~\ref{sec.design} can be directly used with simple extensions.

\vspace{-2mm}
\subsection{Comparison with the MPL Frameworks based on Additive Secret Sharing}
\label{sec.ass}

In the MPL frameworks~\cite{mohassel2017secureml,wagh2019securenn}, such as \texttt{SecureML}~\cite{mohassel2017secureml}, \texttt{SecureNN}~\cite{wagh2019securenn}, based on additive secret sharing~\cite{DBLP:conf/esorics/BogdanovLW08}, the final model can be revealed only when all parties corporate.
Thus, these additive secret sharing based MPL frameworks can meet the first requirement mentioned in Section~\ref{sec.introduction} by setting a sole party to hold all trained shares. However, these additive secret sharing based frameworks cannot meet the second requirement. In these MPL frameworks, once one party drops out, the training will be aborted and must be restarted. Especially, when one party in additive secret sharing based MPL frameworks, e.g. \texttt{SecureML}, intentionally quit the training, the training process cannot be restarted.

In our proposed \pmpl, which is based on vector space secret sharing, the chances of handling the result between the \textit{privileged party} and \textit{assistant parties} are different. Because every authorized set contains the \textit{privileged party} $P_0$, without the participation of $P_0$, \textit{assistant parties} cannot reveal the secret value even if they collude with each other. Moreover, the vector space secret sharing supports multiple ways to reveal results (see Section~\ref{sec:share_reveal} for details), i.e. different linear combinations of \textit{public vectors} held by each party. Therefore, \pmpl can tolerate that one of \textit{assistant parties} drops out.

\vspace{-2mm}
\subsection{Complex Models in MPL Frameworks}
\label{sec.cnn}
\pmpl supports various typical machine learning models, including linear regression, logistic regression, and BP neural networks, following current mainstream MPL frameworks. To further demonstrate the performance of \pmpl, we conduct several experiments on more complex datasets, including Fashion-MNIST and SVHN. We compare the training accuracy of machine learning models trained with \pmpl against the accuracy of machine learning models trained with plaintext data for the 10-class classification.  As is shown in Appendix~\ref{sec.accuracy}, the results show that, under the same model structure, the accuracy of the machine learning models trained with \pmpl is almost the same as that from the training data in plaintext.

For more complex and practical models, i.e. convolutional neural networks (CNN for short), as Max pooling, which is a key component of CNN,  has no efficient secure computation protocol still now, we do not evaluate it in this paper. However, \pmpl now has the potential to support CNN because \pmpl has supported the key components of CNN, including full-connection layer, activation functions, and convolution operation that is essentially matrix multiplication. In future, we will optimize the secure computation protocol of Max pooling to support CNN models.

\vspace{-2mm}
\subsection{Comparison with Federated Learning}
\label{sec.flmpl}
Typical federated learning frameworks~\cite{konevcny2016federated1,konevcny2016federated2} also follow a hierarchical architecture, which has one centralized server and several clients. More specifically, federated learning iteratively executes the three steps as follows:  (1) the centralized server sends the current global model to the clients or a subset of them; (2) each client tunes the global model received from the centralized server with its local data and sends model updates to the centralized server;  (3) the centralized server updates the global model with the local model updates from clients.
In federated learning, each client utilizes its own plaintext data to train a local model, and the communication among parties is coordinated by a centralized server. 

Even though \pmpl and federated learning both follow the hierarchical architecture, the centralized server in federated learning plays a totally different role in the training. It should hold more privileges than the \textit{privileged party} in \pmpl. In \pmpl, the training is performed on shares, and the communication among these parties are in shares too. Thus, no party can infer private information from the intermediate results due to the security guarantees, which is shown in Appendix~\ref{sec.proofs}, of the underlying techniques.  In contrast, in federated learning, the model updates exchanged between clients and the centralized server might contain much sensitive information, which might be leaked~\cite{melis2019exploiting, DBLP:series/lncs/Zhu020} to the centralized server (i.e. the centralized server might get clients' raw data).

\vspace{-2mm}
\subsection{Future Work}
In future, we will optimize the efficiency of \pmpl through reducing the communication rounds of matrix multiplication with truncation and reducing the communication rounds of activation functions evaluation. Meanwhile, we will support more complex machine learning models, such as CNN. 

\vspace{-2mm}
\section{Conclusion}
\label{sec.conclusion}
In this paper, we propose \pmpl, an MPL framework based on the vector space secret sharing. To the best of our knowledge, \pmpl is the first academic work to support a \textit{privileged party} in an MPL framework. \pmpl guarantees that even if two \textit{assistant parties} collude with each other, only the \textit{privileged party} can obtain the final result. Furthermore, \pmpl tolerates one of the two \textit{assistant parties} dropping out during training. That is, \pmpl protects the interests of the \textit{privileged party} while improving the robustness of the framework.
Finally, the experimental results show that the performance of \pmpl is promising when we compare it with state-of-the-art MPL frameworks. Especially,  for the linear regression, \pmpl is $16\times $ faster than \texttt{TF-encrypted} and  $5\times $ for logistic regression in the LAN setting. In the WAN setting, although \pmpl is slower than both \texttt{SecureML} and \texttt{TF-encrypted}, the performance is still promising. Because  \pmpl requires more communication overhead  to ensure both the security (i.e., defending the collusion of two assist parties) and robustness, two requirements from novel practical scenarios.

\vspace{-2mm}
\begin{acks}
This paper is supported by NSFC (No. U1836207, 62172100) and STCSM (No. 21511101600).  We thank all anonymous reviewers for their insightful comments. Weili Han is the corresponding author.
\end{acks}

\bibliographystyle{ACM-Reference-Format}
\bibliography{pmpl}

\appendix

\section{Shares Held by Each Party}

\subsection{Shares During Secure Multiplication }
\label{sec.mulshare}
We show the shares held by each party $P_i$ during the execution of secure multiplication protocol $\prod_{\rm mul} (\mathcal{P},\langle x \rangle, \langle y \rangle)$ (Protocol \ref{pro:mul}) in Table \ref{table.mulshare}.  More specifically, for the first line, each party $P_i$ holds $\langle u \rangle_i, \langle v \rangle_i, \langle h \rangle_i$ by performing $\prod_{\rm vmtgen} (\mathcal{P})$ (Protocol \ref{pro:vmt}) during the offline phase. $P_3$ additionally holds $\langle u \rangle_3, \langle v \rangle_3, \langle h \rangle_3$. The second line in Table \ref{table.mulshare} shows the shares of two inputs $x$ and $y$ held by each party $P_i$. For the rest three lines, they are corresponding to the three steps of $\prod_{\rm mul} (\mathcal{P},\langle x \rangle, \langle y \rangle)$ (Protocol \ref{pro:mul}).

\begin{table*}[t]
\centering
\renewcommand\arraystretch{1}
\caption{Shares held by each party during the execution of $\prod_{\rm mul} (\mathcal{P},\langle x \rangle, \langle y \rangle)$ (Protocol \ref{pro:mul}). For each line, the shares held by each party $P_i$ correspond to each step in $\prod_{\rm mul} (\mathcal{P},\langle x \rangle, \langle y \rangle)$ (Protocol \ref{pro:mul}). }
\label{table.mulshare}
\begin{tabular}{cccc}
\toprule
       Step    & \multicolumn{1}{c}{\textit{Privileged party} $P_0$}                                                                                                                                                                                                                                                                                              & \multicolumn{1}{c}{\textit{Assistant party} $P_1$}                                                                                                                                                     & \textit{Assistant party} $P_2$                                                                                                                                                     \\ \hline
Pre-generating    & \multicolumn{1}{c}{$\langle u \rangle_0$, $\langle u \rangle_3$, $\langle v \rangle_0$, $\langle v \rangle_3$, $\langle h \rangle_0$, $\langle h \rangle_3$}                                                                                                                                                                            & \multicolumn{1}{c}{$\langle u \rangle_1$, $\langle v \rangle_1$, $\langle h \rangle_1$}                                                                                                          & $\langle u \rangle_2$, $\langle v \rangle_2$, $\langle h \rangle_2$ \\ \hline
Inputting    & \multicolumn{1}{c}{$\langle x \rangle_0$, $\langle x \rangle_3$, $\langle y \rangle_0$, $\langle y \rangle_3$}                                                                                                                                                                            & \multicolumn{1}{c}{$\langle x \rangle_1$, $\langle y \rangle_1$}                                                                                                          & $\langle x \rangle_2$, $\langle y \rangle_2$                                                                                                          \\ \hline
Locally computing & \multicolumn{1}{c}{\begin{tabular}[c]{@{}c@{}}$\langle e \rangle_0 = \langle x \rangle_0 + \langle u \rangle_0$\\ $\langle e \rangle_3 = \langle x \rangle_3 + \langle u \rangle_3$\\ $\langle f \rangle_0 =\langle y \rangle_0 + \langle v \rangle_0$\\ $\langle f \rangle_3 =\langle y \rangle_3 + \langle v \rangle_3$\end{tabular}} & \multicolumn{1}{c}{\begin{tabular}[c]{@{}c@{}}$\langle e \rangle_1 = \langle x \rangle_1 + \langle u \rangle_1$\\ $\langle f \rangle_1 =\langle y \rangle_1 + \langle v \rangle_1$\end{tabular}} & \begin{tabular}[c]{@{}c@{}}$\langle e \rangle_2 = \langle x \rangle_2 + \langle u \rangle_2$\\ $\langle f \rangle_2 =\langle y \rangle_2 + \langle v \rangle_2$\end{tabular} \\ \hline
Communicating     & \multicolumn{3}{c}{$\prod_{\rm rec} (\mathcal{P},\langle e \rangle )$ and $\prod_{\rm rec} (\mathcal{P},\langle f \rangle )$}                                                                                                                                                                                                                                                                                                                                                                                                                                                                                                                                                                                       \\ \hline
Locally computing & \multicolumn{1}{c}{\begin{tabular}[c]{@{}c@{}}$\langle z \rangle_0 = \langle x \rangle_0 \cdot f - \langle v \rangle_0 \cdot e+\langle h \rangle_0$\\ $\langle z \rangle_3 = \langle x \rangle_3 \cdot f - \langle v \rangle_3 \cdot e+\langle h \rangle_3$\end{tabular}}                                                                                       & \multicolumn{1}{c}{$\langle z \rangle_1 = \langle x \rangle_1 \cdot f - \langle v \rangle_1 \cdot e+\langle h \rangle_1$}                                                                                    & $\langle z \rangle_2 = \langle x \rangle_2 \cdot f - \langle v \rangle_i \cdot e+\langle h \rangle_2$                                                                                    \\ \bottomrule
\end{tabular}
\end{table*}

\subsection{Shares During Vector Multiplication Triplets Generation}
\label{sec.vmtshare}
We show the shares held by each party $P_i$ during the execution of vector multiplication triplet generation protocol $\prod_{\rm vmtgen} (\mathcal{P})$ (Protocol \ref{pro:vmt}) in Table \ref{table.vmtshare}.  More specifically, the three steps of generating $\langle u \rangle_i$,  $\langle v \rangle_i$ is corresponding to the first three lines of Table \ref{table.vmtshare}. For the four steps of generating $\langle h \rangle_i$, it is corresponding to the last four lines of Table \ref{table.vmtshare}.

\begin{table*}[t]
\centering
\renewcommand\arraystretch{1}
\caption{Shares held by each party during the execution of $\prod_{\rm vmtgen} (\mathcal{P})$ (Protocol \ref{pro:vmt}). For each line, the shares held by each party correspond to each step in $\prod_{\rm vmtgen} (\mathcal{P})$ (Protocol \ref{pro:vmt}). }
\label{table.vmtshare}
\scalebox{0.9}{
\begin{tabular}{cccc}
\toprule
Step                                                                       & \textit{Privileged party} $P_0$                                                                                                                                                                                                                                                                                                                                                                                                     & \textit{Assistant party} $P_1$                                                                                                                                                                                                         & \textit {Assistant party} $P_2$                                                                                                                                                                                                        \\ \hline
Generating random values & two random values $u_0$, $v_0$                                                                                                                                                                                                                                                                                                                                                                                                      & two random values $u_1$, $v_1$                                                                                                                                                                                                      & two random values $u_2$, $v_2$                                                                                                                                                                                                      \\ \hline
\begin{tabular}[c]{@{}c@{}}Executing $\prod_{\rm shr} (P_i,u_i)$ \\ and $\prod_{\rm shr} (P_i,v_i)$\end{tabular} & \begin{tabular}[c]{@{}c@{}}$\langle u_0 \rangle_0, \langle u_1 \rangle_0, \langle u_2 \rangle_0$\\ $\langle v_0 \rangle_0, \langle v_1 \rangle_0, \langle v_2 \rangle_0$\\ $\langle u_0 \rangle_3, \langle u_1 \rangle_3, \langle u_2 \rangle_3$\\ $\langle v_0 \rangle_3, \langle v_1 \rangle_3, \langle v_2 \rangle_3$\end{tabular}                                                                                               & \begin{tabular}[c]{@{}c@{}}$\langle u_0 \rangle_1, \langle u_1 \rangle_1, \langle u_2 \rangle_1$ \\ $\langle v_0 \rangle_1, \langle v_1 \rangle_1, \langle v_2 \rangle_1$\end{tabular}                                              & \begin{tabular}[c]{@{}c@{}}$\langle u_0 \rangle_2, \langle u_1 \rangle_2, \langle u_2 \rangle_2$\\ $\langle v_0 \rangle_2, \langle v_1 \rangle_2, \langle v_2 \rangle_2$\end{tabular}                                               \\ \hline
Locally computing                                                             & \begin{tabular}[c]{@{}c@{}}$\langle u \rangle_0 = \langle u_0 \rangle_0 + \langle u_1 \rangle_0 + \langle u_2 \rangle_0$\\ $\langle v \rangle_0 = \langle v_0 \rangle_0 + \langle v_1 \rangle_0 + \langle v_2 \rangle_0$\\ $\langle u \rangle_3 = \langle u_0 \rangle_ 3+\langle u_1 \rangle_3 + \langle u_2 \rangle_3$\\ $\langle v \rangle_3 = \langle v_0 \rangle_3 +\langle v_1 \rangle_3 + \langle v_2 \rangle_3$\end{tabular} & \begin{tabular}[c]{@{}c@{}}$\langle u \rangle_1 = \langle u_0 \rangle_1 +\langle u_1 \rangle_1 + \langle u_2 \rangle_1$\\ $\langle v \rangle_1 = \langle v_0 \rangle_1 +\langle v_1 \rangle_1 + \langle v_2 \rangle_1$\end{tabular} & \begin{tabular}[c]{@{}c@{}}$\langle u \rangle_2 = \langle u_0 \rangle_2 +\langle u_1 \rangle_2 + \langle u_2 \rangle_2$\\ $\langle v \rangle_2 = \langle v_0 \rangle_2 +\langle v_1 \rangle_2 + \langle v_2 \rangle_2$\end{tabular} \\ \hline
Secure computing                                                              & \begin{tabular}[c]{@{}c@{}}$[ u_{0}*v_{1}+v_{0}*u_{1} ]_0$\\ $[ u_{0}*v_{2}+v_{0}*u_{2}]_0$\end{tabular}                                                                                                                                                                                                                                                                                                    & \begin{tabular}[c]{@{}c@{}}$[ u_{0}*v_{1}+v_{0}*u_{1} ]_1$\\ $[ u_{1}*v_{2}+v_{1}*u_{2} ]_1$\end{tabular}                                                                                                   & \begin{tabular}[c]{@{}c@{}}$[ u_{0}*v_{2}+v_{0}*u_{2} ]_2$\\ $[ u_{1}*v_{2}+v_{1}*u_{2} ]_2$\end{tabular}                                                                                                   \\ \hline
Locally computing    & \begin{tabular}[c]{@{}c@{}}$h_{0}= u_{0}*v_{0}+[ u_{0}*v_{1}+v_{0}*u_{1} ]_0$\\ $+[ u_{0}*v_{2}+v_{0}*u_{2}]_0$\end{tabular}                                                                                                                                                                                                                                                                                & \begin{tabular}[c]{@{}c@{}}$h_{1}=u_{1}*v_{1}+[ u_{0}*v_{1}+v_{0}*u_{1} ]_1$\\ $+[ u_{1}*v_{2}+v_{1}*u_{2} ]_1$\end{tabular}                                                                                & \begin{tabular}[c]{@{}c@{}}$h_{2} = u_{2}*v_{2}+[ u_{0}*v_{2}+v_{0}*u_{2} ]_2$\\ $+[ u_{1}*v_{2}+v_{1}*u_{2} ]_2$\end{tabular}                                                                              \\ \hline
Executing $\prod_{\rm shr} (P_i,h_i)$                                                           & \begin{tabular}[c]{@{}c@{}}$\langle h_0 \rangle_0, \langle h_1 \rangle_0, \langle h_2 \rangle_0$\\ $\langle h_0 \rangle_3, \langle h_1 \rangle_3, \langle h_2 \rangle_3$\end{tabular}                                                                                                                                                                                                                                               & $\langle h_0 \rangle_1, \langle h_1 \rangle_1, \langle h_2 \rangle_1$                                                                                                                                                               & $\langle h_0 \rangle_2, \langle h_1 \rangle_2, \langle h_2 \rangle_2$                                                                                                                                                               \\ \hline
Locally computing                                                             & \begin{tabular}[c]{@{}c@{}}$\langle h \rangle_0 = \langle h_0 \rangle_0 + \langle h_1 \rangle_0 + \langle h_2 \rangle_0$\\ $\langle h \rangle_3 = \langle h_0 \rangle_3 + \langle h_1 \rangle_3 + \langle h_2 \rangle_3$\end{tabular}                                                                                                                                                                                               & $\langle h \rangle_1 = \langle h_0 \rangle_1 +\langle h_1 \rangle_1 + \langle h_2 \rangle_1$                                                                                                                                        & $\langle h \rangle_2 = \langle h_0 \rangle_2 +\langle h_1 \rangle_2 + \langle h_2 \rangle_2$                                                                                                                                        \\ \bottomrule
\end{tabular}
}
\end{table*}

\section{Security of Our Designs}
\label{sec.proofs}
In this section, we introduce the security of our design using the standard real/ideal world paradigm.  We use $\mathcal{S}$ to denote an ideal-world static adversary (simulator) for a real-world adversary
. $\mathcal{S}$ acts as the honest parties and simulates the messages received by real-world adversary during the protocol. For each of the constructions, we provide the simulation proof for the case of corrupt of $P_0$ and the case of corrupt $P_1$ and $P_2$ (i.e. $P_1$ and $P_2$ collude with each other).

\begin{figure}[ht]
    \centering
    \fbox{
    \parbox{0.4\textwidth}{
        \smallskip \textbf{Functionality} $\mathcal{F}_{\rm shr}$ \\
        \smallskip \textbf{Input:}  
        \begin{itemize}
            \item  $P_0$ inputs $x$.
        \end{itemize}
        \smallskip \textbf{Output:} 
        \begin{itemize}
            \item $P_0$ outputs $\langle x \rangle_0$ and $\langle x \rangle _3$;
            \item $P_1$ outputs $\langle x \rangle_1$;
            \item $P_2$ outputs $\langle x \rangle_2$.
        \end{itemize}
    }
    }
    \caption{Functionality $\mathcal{F}_{\rm shr}$}
    \label{funct:share}
\end{figure}

\smallskip
\noindent\textbf{Sharing Protocol:}
The ideal functionality $\mathcal{F}_{\rm shr}$ realising sharing protocol $\prod_{\rm shr} (P_i,x)$ (Protocol \ref{pro:ss}) is presented in Figure \ref{funct:share}. Here we assume that $P_0$ inputs $x$.

\begin{theorem}
Sharing protocol $\prod_{\rm shr} (P_i,x)$ (Protocol \ref{pro:ss}) securely realizes the functionality $\mathcal{F}_{\rm shr}$ (Functionality \ref{funct:share}) in the presence of static semi-honest adversary.

Proof: We present the simulation for the case for corrupt $P_0$ and the case for corrupt $P_1$ and $P_2$ as shown in Figure~\ref{simulator:share0} and Figure~\ref{simulator:share1} respectively. 
\end{theorem}

\begin{figure}[ht]
    \centering
    \vspace{-4mm}
    \fbox{
    \parbox{0.4\textwidth}{
        \smallskip 
        \textbf{Simulator} $\mathcal{S}_{\rm shr}^{\rm P_0}$ 
        \begin{algorithmic}[1]
            \STATE $\mathcal{S}_{\rm shr}^{\rm P_0}$ receives $x$ and $\varPhi (\mathcal{P})$ from $P_0$. 
            \STATE $\mathcal{S}_{\rm shr}^{\rm P_0}$ selects two random values $s_1,s_2$, and constructs a vector $\vec{s} = (x,s_1,s_2)^T$. 
            \STATE $\mathcal{S}_{\rm shr}^{\rm P_0}$ computes
                \begin{equation*}
                    \begin{aligned}
                        & \langle x \rangle_0 = \varPhi(0) \times \vec{s}, ~ &\langle x \rangle_1 = \varPhi(1) \times \vec{s} \\
                        &\langle x \rangle_2 = \varPhi(2) \times \vec{s}, ~ &\langle x \rangle_3 = \varPhi(3) \times \vec{s} 
                    \end{aligned}
                \end{equation*}
            \STATE $\mathcal{S}_{\rm shr}^{\rm P_0}$ outputs $(x,\langle x \rangle_0,\langle x \rangle_1,\langle x \rangle_2,\langle x \rangle_3).$
        \end{algorithmic}
        }
    }
    \vspace{-3mm}
    \caption{Simulator $\mathcal{S}_{\rm shr}^{\rm P_0}$}
    \label{simulator:share0}
\vspace{-3mm}
\end{figure}

\begin{figure}[ht]
    \centering
    \vspace{-3mm}
    \fbox{
    \parbox{0.4\textwidth}{
        \smallskip 
        \textbf{Simulator} $\mathcal{S}_{\rm shr}^{\rm P_1,P_2}$ 
        \begin{algorithmic}[1]
            \STATE $\mathcal{S}_{\rm shr}^{\rm P_1,P_2}$ receives $\varPhi (\mathcal{P})$ from $P_1$. 
            \smallskip
            \STATE $\mathcal{S}_{\rm shr}^{\rm P_1,P_2}$ selects three random values $x,s_1,s_2$, and constructs a vector $\vec{s} = (x,s_1,s_2)^T$.
            \smallskip
            \STATE $\mathcal{S}_{\rm shr}^{\rm P_1,P_2}$ computes
                \begin{equation*}
                    \langle x \rangle_1 = \varPhi(1) \times \vec{s}, ~ \langle x \rangle_2 = \varPhi(2) \times \vec{s}
                \end{equation*}
            \STATE $\mathcal{S}_{\rm shr}^{\rm P_1,P_2}$ outputs $(\langle x \rangle_1,\langle x \rangle_2).$
        \end{algorithmic}
        }
    }
    \vspace{-3mm}
    \caption{Simulator $\mathcal{S}_{\rm shr}^{\rm P_1,P_2}$}
    \label{simulator:share1}
    \vspace{-3mm}
\end{figure}

We denote $\textbf{view}_{P_0}^{shr}$ and $\textbf{view}_{P_1,P_2}^{shr}$ as the views of $P_0$ and $P_1,P_2$ respectively. We note that $P_0$'s view and $\mathcal{S}_{\rm shr}^{\rm P_0}$'s output are identical, the probability distribution of $P_1$ and $P_2$'s views and $\mathcal{S}_{\rm shr}^{\rm P_1,P_2}$'s output are identical. Therefore we have the following equations:

\begin{equation*}
\begin{aligned}
&\mathcal{S}_{\rm shr}^{\rm P_0}(x,\langle x \rangle_0,\langle x \rangle_3) \cong \textbf{view}_{P_0}^{shr}(x,\langle x \rangle_k,k \in\{0,1,2,3\}) \\
&\mathcal{S}_{\rm shr}^{\rm P_1,P_2}(\emptyset,\langle x \rangle_1,\langle x \rangle_2) \cong \textbf{view}_{P_1,P_2}^{shr}(x,\langle x \rangle_k,k \in\{0,1,2,3\})
\end{aligned}
\end{equation*}

\begin{figure}[ht]
    \centering
    \vspace{-3mm}
    \fbox{
    \parbox{0.4\textwidth}{
        \smallskip \textbf{Functionality} $\mathcal{F}_{\rm rec}$ \\
        \smallskip \textbf{Input:}  
        \begin{itemize}
            \item $P_0$ inputs $\langle x \rangle_0$;
            \item $P_1$ inputs $\langle x \rangle_1$;
            \item $P_2$ inputs $\langle x \rangle_2$.
        \end{itemize}

        \smallskip \textbf{Output:}
        \begin{itemize}
            \item $P_0$, $P_1$ and $P_2$ all output $x$. 
        \end{itemize}
        
        }
    }
    \vspace{-3mm}
    \caption{Functionality $\mathcal{F}_{\rm rec}$}
    \label{funct:rec}
    \vspace{-3mm}
\end{figure}

\smallskip
\noindent\textbf{Reconstruction Protocol:}
The ideal functionality $\mathcal{F}_{\rm rec}$ realising reconstruction protocol $\prod_{\rm rec} (\mathcal{P},\langle x \rangle )$ (Protocol \ref{pro:rec}) is presented in Figure \ref{funct:rec}. Here, we only consider the case of no party drops out.

\begin{theorem}
Reconstruction protocol $\prod_{\rm rec} (P_i,\langle x \rangle)$ (Protocol \ref{pro:rec}) securely realizes the functionality $\mathcal{F}_{\rm rec}$ (Figure~\ref{funct:rec}) in the presence of static semi-honest adversary.

Proof: We present the simulation for the case for corrupt $P_0$ and the case for corrupt $P_1$ and $P_2$ as shown in Figure~\ref{simulator:rec0} and Figure~\ref{simulator:rec1} respectively.
\end{theorem}

We denote $\textbf{view}_{P_0}^{rec}$ and $\textbf{view}_{P_1,P_2}^{rec}$ as the views of $P_0$ and $P_1,P_2$ respectively. We note that the probability distribution of $P_0$'s view and $\mathcal{S}_{\rm rec}^{\rm P_0}$'s output are identical, the probability distribution of $P_1$ and $P_2$'s views and $\mathcal{S}_{\rm rec}^{\rm P_1,P_2}$'s output are identical. Therefore we have the following equations:

\begin{equation*}
\begin{aligned}
&\mathcal{S}_{\rm rec}^{\rm P_0}(\langle x \rangle_0, x) \cong \textbf{view}_{P_0}^{rec}(\langle x \rangle_0,\langle x \rangle_1,\langle x \rangle_2, x)\\
&\mathcal{S}_{\rm rec}^{\rm P_1,P_2}(\langle x \rangle_1,\langle x \rangle_2, x) \cong \textbf{view}_{P_1,P_2}^{rec}(\langle x \rangle_0,\langle x \rangle_1,\langle x \rangle_2, x)
\end{aligned}
\end{equation*}

\begin{figure}[ht]
    \centering
    \vspace{-3mm}
    \fbox{
    \parbox{0.4\textwidth}{
        \smallskip \textbf{Simulator} $\mathcal{S}_{\rm rec}^{\rm P_0}$
        \begin{algorithmic}[1]
        \STATE $\mathcal{S}_{\rm rec}^{\rm P_0}$ receives $\langle x \rangle_0$ and $c_0,c_1,c_2$ from $P_0$. 
        \smallskip
        \STATE $\mathcal{S}_{\rm rec}^{\rm P_0}$ selects two random values $\langle x \rangle_1,\langle x \rangle_2$. 
        \smallskip
        \STATE $\mathcal{S}_{\rm rec}^{\rm P_0}$ computes
        \begin{equation*}
                x = c_0\cdot \langle x \rangle_0 + c_1 \cdot \langle x \rangle_1 + c_2 \cdot \langle x \rangle_2
        \end{equation*}
        \STATE $\mathcal{S}_{\rm rec}^{\rm P_0}$ outputs $(\langle x \rangle_0,\langle x \rangle_1,\langle x \rangle_2, x).$
        \end{algorithmic}
        }
    }
    \vspace{-3mm}
    \caption{Simulator $\mathcal{S}_{\rm rec}^{\rm P_1,P_2}$}
    \label{simulator:rec0}
\end{figure}

\begin{figure}[ht]
    \centering
    \fbox{
    \parbox{0.4\textwidth}{
        \smallskip \textbf{Simulator} $\mathcal{S}_{\rm rec}^{\rm P_1,P_2}$
        \begin{algorithmic}[1]
            \STATE $\mathcal{S}_{\rm rec}^{\rm P_1,P_2}$ receives  $\langle x \rangle_1, \langle x \rangle_2$ and $c_0,c_1,c_2$ from $P_1,P_2$.
            \STATE $\mathcal{S}_{\rm rec}^{\rm P_1,P_2}$ selects a random value $\langle x \rangle_0$.
            \STATE $\mathcal{S}_{\rm rec}^{\rm P_1,P_2}$ computes
            \begin{equation*}
                \begin{aligned}
                    x &= c_0\cdot \langle x \rangle_0 + c_1 \cdot \langle x \rangle_1 + c_2 \cdot \langle x \rangle_2
                \end{aligned}
            \end{equation*}
            \STATE $\mathcal{S}_{\rm rec}^{\rm P_1,P_2}$ outputs $(\langle x \rangle_0, \langle x \rangle_1, \langle x \rangle_2, x).$
        \end{algorithmic}
        }
    }
    \vspace{-3mm}
    \caption{Simulator $\mathcal{S}_{\rm rec}^{\rm P_1,P_2}$}
    \label{simulator:rec1}
\end{figure}

\smallskip
\noindent\textbf{Multiplication Protocol:}
The ideal functionality $\mathcal{F}_{\rm mul}$ realising multiplication protocol $\prod_{\rm mul} (\mathcal{P},\langle x \rangle, \langle y \rangle)$ (Protocol \ref{pro:mul}) is presented in Figure \ref{funct:mul}.

\begin{figure}[ht]
    \centering
    \vspace{-3mm}
    \fbox{
    \parbox{0.4\textwidth}{
        \smallskip \textbf{Functionality} $\mathcal{F}_{\rm mul}$ \\
        \smallskip \textbf{Input:} 
        \begin{itemize}
            \item $P_0$ inputs $\langle x \rangle_0, \langle y \rangle_0$ and $\langle x \rangle_3, \langle y \rangle_3$;
            \item $P_1$ inputs $\langle x \rangle_1, \langle y \rangle_1$;
            \item $P_2$ inputs $\langle x \rangle_2, \langle y \rangle_2$.
        \end{itemize}
        \smallskip \textbf{Output:} 
        \begin{itemize}
            \item $P_0$ outputs $\langle z \rangle_0$ and $\langle z \rangle_1$;
            \item $P_1$ outputs $\langle z \rangle_1$;
            \item $P_2$ outputs $\langle z \rangle_2$, where $z=x\cdot y$.
        \end{itemize}
          
    }
    }
    \vspace{-3mm}
    \caption{Functionality $\mathcal{F}_{\rm mul}$}
    \label{funct:mul}
    \vspace{-3mm}
\end{figure}

\begin{theorem}
Multiplication protocol $\prod_{\rm mul} (\mathcal{P},\langle x \rangle, \langle y \rangle)$(Protocol \ref{pro:mul}) securely realizes the functionality $\mathcal{F}_{\rm mul}$ (Figure \ref{funct:mul}) in the presence of static semi-honest adversary.

Proof: We present the simulation for the case for corrupt $P_0$ and the case for corrupt $P_1$ and $P_2$ as shown in Figure~\ref{simulator:mul0} and Figure~\ref{simulator:mul1} respectively.
\end{theorem}

\begin{figure}[ht]
    \centering
    \vspace{-3mm}
    \fbox{
    \parbox{0.4\textwidth}{
        \smallskip \textbf{Simulator} $\mathcal{S}_{\rm mul}^{\rm P_0}$
        \begin{algorithmic}[1]
            \STATE $\mathcal{S}_{\rm mul}^{\rm P_0}$ receives $\langle x \rangle_0,\langle y \rangle_0,\langle x \rangle_3,\langle y \rangle_3$ from $P_0$.
            \STATE $\mathcal{S}_{\rm mul}^{\rm P_0}$ receives $\langle u \rangle_0,\langle v \rangle_0,\langle h \rangle_0,\langle u \rangle_3,\langle v \rangle_3,\langle h \rangle_3$ from $P_0$.
            \STATE $\mathcal{S}_{\rm mul}^{\rm P_0}$ computes
            \begin{equation*}
                \begin{aligned}
                    & \langle e \rangle_0 = \langle x \rangle_0+\langle u \rangle_0, ~ &\langle f \rangle_0 = \langle y \rangle_0+\langle v \rangle_0 \\
                    & \langle e \rangle_3 = \langle x \rangle_3+\langle u \rangle_3, ~ &\langle f \rangle_3 = \langle y \rangle_3+\langle v \rangle_3
                \end{aligned}
            \end{equation*}
            \STATE $\mathcal{S}_{\rm mul}^{\rm P_0}$ selects random values $\langle e \rangle_1,\langle f \rangle_1,\langle e \rangle_2,\langle f \rangle_2$.
            \STATE $\mathcal{S}_{\rm mul}^{\rm P_0}$ computes
            \begin{equation*}
            \begin{aligned}
                  e &= c_0\cdot \langle e \rangle_0 + c_1 \cdot \langle e \rangle_1 + c_2 \cdot \langle e \rangle_2  \\
                  f &= c_0\cdot \langle f \rangle_0 + c_1 \cdot \langle f \rangle_1 + c_2 \cdot \langle f \rangle_2 
            \end{aligned}
            \end{equation*}
            \STATE $\mathcal{S}_{\rm mul}^{\rm P_0}$ computes
            \begin{equation*}
                \begin{aligned}
                    \langle z \rangle_0 &= \langle x \rangle_0 \cdot f - \langle v \rangle_0 \cdot e + \langle h \rangle_0\\
                    \langle z \rangle_3 &= \langle x \rangle_3 \cdot f - \langle v \rangle_3 \cdot e + \langle h \rangle_3
                \end{aligned}
            \end{equation*}
            \STATE $\mathcal{S}_{\rm mul}^{\rm P_0}$ outputs $(\langle x \rangle_0,\langle x \rangle_3,\langle e \rangle_j,\langle f \rangle_j,\langle z \rangle_0,\langle z \rangle_3,j\in \{1,2\})$.
        \end{algorithmic}
        }
    }
    \vspace{-3mm}
    \caption{Simulator $\mathcal{S}_{\rm mul}^{\rm P_0}$}
    \label{simulator:mul0}
    \vspace{-3mm}
\end{figure}

\begin{figure}[ht]
    \centering
    \fbox{
    \parbox{0.4\textwidth}{
        \smallskip \textbf{Simulator}$\mathcal{S}_{\rm mul}^{\rm P_1,P_2}$
        \begin{algorithmic}[1]
            \STATE $\mathcal{S}_{\rm mul}^{\rm P_1,P_2}$ receives $\langle x \rangle_1,\langle y \rangle_1,\langle x \rangle_2,\langle y \rangle_2$ from $P_1,P_2$. 
            \STATE $\mathcal{S}_{\rm mul}^{\rm P_1,P_2}$ receives $\langle u \rangle_1,\langle v \rangle_1,\langle h \rangle_1,\langle u \rangle_2,\langle v \rangle_2,\langle h \rangle_2$ from $P_1,P_2$.
            \STATE $\mathcal{S}_{\rm mul}^{\rm P_1,P_2}$ computes
            \begin{equation*}
                \begin{aligned}
                    & \langle e \rangle_1 = \langle x \rangle_1+\langle u \rangle_1, ~ &\langle f \rangle_1 = \langle y \rangle_1+\langle v \rangle_1 \\
                    & \langle e \rangle_2 = \langle x \rangle_2+\langle u \rangle_2, ~ &\langle f \rangle_2 = \langle y \rangle_2+\langle v \rangle_2
                \end{aligned}
            \end{equation*}
            \STATE $\mathcal{S}_{\rm mul}^{\rm P_1,P_2}$ selects random values $\langle e \rangle_0,\langle f \rangle_0$.
            \STATE $\mathcal{S}_{\rm mul}^{\rm P_1,P_2}$ computes
            \begin{equation*}
                \begin{aligned}
                    e &= c_0\cdot \langle e \rangle_0 + c_1 \cdot \langle e \rangle_1 + c_2 \cdot \langle e \rangle_2 \\
                    f &= c_0\cdot \langle f \rangle_0 + c_1 \cdot \langle f \rangle_1 + c_2 \cdot \langle f \rangle_2 
                \end{aligned}
            \end{equation*}
            \STATE $\mathcal{S}_{\rm mul}^{\rm P_1,P_2}$ computes
            \begin{equation*}
                \begin{aligned}
                    \langle z \rangle_1 &= \langle x \rangle_1 \cdot f - \langle v \rangle_1 \cdot e + \langle h \rangle_1\\
                    \langle z \rangle_2 &= \langle x \rangle_2 \cdot f - \langle v \rangle_2 \cdot e + \langle h \rangle_2
                \end{aligned}
            \end{equation*}
            \STATE $\mathcal{S}_{\rm mul}^{\rm P_1,P_2}$ outputs $(\langle x \rangle_j,\langle e \rangle_0,\langle f \rangle_0,\langle z \rangle_j,j\in \{1,2\})$.
        \end{algorithmic}
        }
    }
    \vspace{-3mm}
    \caption{Simulator $\mathcal{S}_{\rm mul}^{\rm P_1,P_2}$}
    \label{simulator:mul1}
    \vspace{-3mm}
\end{figure}

We denote $\textbf{view}_{P_0}^{mul}$ and $\textbf{view}_{P_1,P_2}^{mul}$ as the views of $P_0$ and $P_1,P_2$ respectively. We note that the probability distribution of $P_0$'s view and $\mathcal{S}_{\rm mul}^{\rm P_0}$'s output are identical, $P_1$ and $P_2$'s view and $\mathcal{S}_{\rm shr}^{\rm P_1,P_2}$'s output are identical. Therefore we have the following equations:

\begin{equation*}
\begin{aligned}
  \mathcal{S}_{\rm mul}^{\rm P_0}(\langle x \rangle_0,\langle y \rangle_0,\langle x \rangle_3,\langle y \rangle_3,\langle z \rangle_0,\langle z \rangle_3) \cong \\ \textbf{view}_{P_0}^{mul}(\langle x \rangle_k,\langle y \rangle_k,\langle z \rangle_k,k\in\{0,1,2,3\}) \\
  \mathcal{S}_{\rm mul}^{\rm P_1,P_2}(\langle x \rangle_1,\langle y \rangle_1,\langle x \rangle_2,\langle y \rangle_2,\langle z \rangle_1,\langle z \rangle_2) \cong \\ \textbf{view}_{P_1,P_2}^{mul}(\langle x \rangle_k,\langle y \rangle_k,\langle z \rangle_k,k\in\{0,1,2,3\})
\end{aligned}
\end{equation*}

\smallskip
\noindent\textbf{Sharing conversion Protocol:}
Here, we only analyze the security of  protocol $\prod_{\rm a2v} (\mathcal{P},[ x ] )$ (Protocol \ref{pro:a2v}) since protocol $\prod_{\rm v2a} (\mathcal{P},\langle x \rangle )$ is executed locally. 
The ideal functionality $\mathcal{F}_{\rm a2v}$ realising protocol $\prod_{\rm a2v} (\mathcal{P},[ x ])$ (Protocol \ref{pro:a2v}) is presented in Figure \ref{funct:a2v}.

\begin{figure}[ht]
    \centering
    \vspace{-3mm}
    \fbox{
    \parbox{0.4\textwidth}{
        \smallskip \textbf{Functionality} $\mathcal{F}_{\rm a2v}$ \\
        \smallskip \textbf{Input:}
        \begin{itemize}
            \item $P_0$ inputs $[x]_0$;
            \item $P_1$ inputs $[x]_1$;
            \item $P_2$ inputs $[x]_2$.
        \end{itemize}
        \smallskip \textbf{Output:}
        \begin{itemize}
            \item $P_0$ outputs $\langle x \rangle_0$ and $\langle x \rangle_3$;
            \item $P_1$ outputs $\langle x \rangle_1$;
            \item $P_2$ outputs $\langle x \rangle_2$.
        \end{itemize}
    }
    }
    \vspace{-3mm}
    \caption{Functionality $\mathcal{F}_{\rm a2v}$}
    \label{funct:a2v}
\end{figure}

\begin{theorem}
Sharing conversion protocol $\prod_{\rm a2v} (\mathcal{P},[ x ]$ (Protocol \ref{pro:a2v}) securely realizes the functionality $\mathcal{F}_{\rm a2v}$ (Figure~\ref{funct:a2v}) in the presence of static semi-honest adversary.

Proof: We present the simulation for the case for corrupt $P_0$ and the case for corrupt $P_1$ and $P_2$ as shown in Figure~\ref{simulator:a2v0} and Figure~\ref{simulator:a2v1} respectively.
\end{theorem}

\begin{figure}[ht]
    \centering
    \vspace{-3mm}
    \fbox{
    \parbox{0.4\textwidth}{
        \smallskip \textbf{Simulator}$\mathcal{S}_{\rm a2v}^{\rm P_0}$
        \begin{algorithmic}[1]
            \STATE $\mathcal{S}_{\rm a2v}^{\rm P_0}$ receives $[x]_0,[x]_3$ and $c_0,a_1,a_2,a_3,\langle k \rangle_0$ from $P_0$.
            \STATE $\mathcal{S}_{\rm a2v}^{\rm P_0}$ selects random values $\langle x+k \rangle_1$,$\langle x+k \rangle_2$,
            \STATE $\mathcal{S}_{\rm a2v}^{\rm P_0}$ computes 
            \begin{equation*}
            \begin{aligned}
                 \langle x \rangle_0 &= [x]_0 /c_0\\
                 \langle x \rangle_3 &= a_1 \cdot \langle x + k \rangle_1+a_2 \cdot \langle x + k \rangle_2 - \langle k \rangle_3
            \end{aligned}
            \end{equation*}
            \STATE $\mathcal{S}_{\rm a2v}^{\rm P_0}$ outputs $([x]_0$, $[x]_3$, $\langle x+k \rangle_1$, $\langle x+k \rangle_2$, $\langle x \rangle_0$, $\langle x \rangle_3)$.
        \end{algorithmic}
        }
    }
    \vspace{-3mm}
    \caption{Simulator $\mathcal{S}_{\rm a2v}^{\rm P_0}$}
    \label{simulator:a2v0}
\end{figure}

\begin{figure}[ht]
    \centering
    \vspace{-3mm}
    \fbox{
    \parbox{0.4\textwidth}{
        \smallskip \textbf{Simulator}$\mathcal{S}_{\rm a2v}^{\rm P_1,P_2}$
        \begin{algorithmic}[1]
            \STATE $\mathcal{S}_{\rm a2v}^{\rm P_1,P_2}$ receives $[x]_1,[x]_2$ and $c_1,c_2,\langle k \rangle_1,\langle k \rangle_2$ from $P_1,P_2$.
            \STATE $\mathcal{S}_{\rm a2v}^{\rm P_1,P_2}$ computes 
            \begin{equation*}
            \begin{aligned}
                 &\langle x \rangle_1 = [x]_1 /c_1,~ \langle x \rangle_2 = [x]_2 /c_2\\
                 &\langle x+k \rangle_1=\langle x \rangle_1+\langle k \rangle_1,~\langle x+k \rangle_2=\langle x \rangle_2+\langle k \rangle_2
            \end{aligned}
            \end{equation*}
            \STATE $\mathcal{S}_{\rm a2v}^{\rm P_1,P_2}$ outputs $([x]_1, [x]_2, \langle x \rangle_1, \langle x \rangle_2)$.
        \end{algorithmic}
        }
    }
    \vspace{-3mm}
    \caption{Simulator $\mathcal{S}_{\rm a2v}^{\rm P_1,P_2}$}
    \label{simulator:a2v1}
\end{figure}

We denote $\textbf{view}_{P_0}^{a2v}$ and $\textbf{view}_{P_1,P_2}^{a2v}$ as the views of $P_0$ and $P_1,P_2$ respectively. We note that the probability distribution of $P_0$'s view and $\mathcal{S}_{\rm a2v}^{\rm P_0}$'s output are identical, $P_1$ and $P_2$'s view and $\mathcal{S}_{\rm a2v}^{\rm P_1,P_2}$'s output are identical. Therefore we have the following equations:

\begin{small}
\begin{equation*}
\begin{aligned}
  &\mathcal{S}_{\rm a2v}^{\rm P_0}([x]_0,[x]_3,\langle x \rangle_0,\langle x \rangle_3) \cong \textbf{view}_{P_0}^{a2v}([x]_i,\langle x \rangle_k,k\in\{0,1,2,3\}) \\
 &\mathcal{S}_{\rm a2v}^{\rm P_1,P_2}([x]_1,[x]_2,\langle x \rangle_1,\langle x \rangle_2) \cong \textbf{view}_{P_1,P_2}^{a2v}([x]_i,\langle x \rangle_k,k\in\{0,1,2,3\})
\end{aligned}
\end{equation*}
\end{small}

\noindent\textbf{Truncation Protocol:}
The ideal functionality $\mathcal{F}_{\rm trunc}$ realizing truncation protocol $\prod_{\rm trunc} (\mathcal{P},\langle z \rangle)$ (Protocol \ref{pro:trunc}) is presented in Figure \ref{funct:trunc}.

\begin{figure}[ht]
    \centering
    \vspace{-3mm}
    \fbox{
    \parbox{0.4\textwidth}{
        \smallskip \textbf{Functionality} $\mathcal{F}_{\rm trunc}$ \\
        \smallskip \textbf{Input:}  
        \begin{itemize}
            \item $P_0$ inputs $\langle z \rangle_0$;
            \item $P_1$ inputs $\langle z \rangle_1$;
            \item $P_2$ inputs $\langle z \rangle_2$.
        \end{itemize}
        \smallskip \textbf{Output:}  
        \begin{itemize}
            \item $P_0$ outputs $\langle z' \rangle_0$ and $\langle z' \rangle_3$;
            \item $P_1$ outputs $\langle z' \rangle_1$;
            \item $P_2$ outputs $\langle z' \rangle_2$, where $z' = z/ 2^{\ell_f}$.
        \end{itemize}
    }
    }
    \vspace{-3mm}
    \caption{Functionality $\mathcal{F}_{\rm trunc}$}
    \label{funct:trunc}
    \vspace{-3mm}
\end{figure}

\begin{theorem}
Truncation protocol $\prod_{\rm trunc} (\mathcal{P},\langle z \rangle$ (Protocol \ref{pro:trunc}) securely realizes the functionality $\mathcal{F}_{\rm trunc}$ (Functionality \ref{funct:trunc}) in the presence of static semi-honest adversary.

Proof: We present the simulation for the case for corrupt $P_0$ and the case for corrupt $P_1$ and $P_2$ as shown in Figure~\ref{simulator:trunc0} and Figure~\ref{simulator:trunc1} respectively.
\end{theorem}

\begin{figure}[ht]
    \centering
    \fbox{
    \parbox{0.4\textwidth}{
        \smallskip \textbf{Simulator} $\mathcal{S}_{\rm trunc}^{\rm P_0}$
        \begin{algorithmic}[1]
            \STATE $\mathcal{S}_{\rm trunc}^{\rm P_0}$ receives $\langle z \rangle_0$ and $c_0,c_1,c_2,\langle r \rangle_0,\langle r' \rangle_0,\langle r' \rangle_3$ from $P_0$.
            \STATE $\mathcal{S}_{\rm trunc}^{\rm P_0}$ selects random values $\langle z-r \rangle_1,\langle z-r \rangle_2$.
            \STATE $\mathcal{S}_{\rm trunc}^{\rm P_0}$ computes
            \begin{equation*}
                \begin{aligned}
                    z-r &= c_0\cdot (\langle z \rangle_0-\langle r \rangle_0) + c_1 \cdot \langle z-r \rangle_1 + c_2 \cdot \langle z-r \rangle_2 \\
                    \langle z' \rangle_0 &= (z-r)/(2^{\ell_f} \cdot c_0) + \langle r' \rangle_0\\
                    \langle z' \rangle_3 &= \langle r' \rangle_3
                \end{aligned}
            \end{equation*}
            \STATE $\mathcal{S}_{\rm trunc}^{\rm P_0}$ outputs $(\langle z \rangle_0, \langle z-r \rangle_1, \langle z-r \rangle_2, \langle z' \rangle_0,\langle z' \rangle_3)$.
        \end{algorithmic}
        }
    }
    \vspace{-3mm}
    \caption{Simulator $\mathcal{S}_{\rm trunc}^{\rm P_0}$}
    \label{simulator:trunc0}
\end{figure}

\begin{figure}[h]
    \centering
    \fbox{
    \parbox{0.4\textwidth}{
        \smallskip \textbf{Simulator}$\mathcal{S}_{\rm trunc}^{\rm P_1,P_2}$
        \begin{algorithmic}[1]
            \STATE $\mathcal{S}_{\rm trunc}^{\rm P_1,P_2}$ receives $\langle r' \rangle_1,\langle r' \rangle_2$ from $P_1,P_2$.
            \STATE $\mathcal{S}_{\rm trunc}^{\rm P_1,P_2}$ computes
            \begin{equation*}
            \begin{aligned}
                &\langle z-r \rangle_1 =\langle z \rangle_1 -\langle r \rangle_1~
              \langle z-r \rangle_2 =\langle z \rangle_2 -\langle r \rangle_2\\
                &\langle z' \rangle_1 = \langle r' \rangle_1~
              \langle z' \rangle_2 = \langle r' \rangle_2
            \end{aligned}
            \end{equation*}
            \STATE $\mathcal{S}_{\rm trunc}^{\rm P_1,P_2}$ outputs $(\langle z \rangle_1, \langle z \rangle_2, \langle z' \rangle_1,\langle z' \rangle_2)$.
        \end{algorithmic}
        }
    }
    \vspace{-3mm}
    \caption{Simulator $\mathcal{S}_{\rm trunc}^{\rm P_1,P_2}$}
    \label{simulator:trunc1}
    \vspace{-3mm}
\end{figure}

We denote $\textbf{view}_{P_0}^{trunc}$ and $\textbf{view}_{P_1,P_2}^{trunc}$ as the views of $P_0$ and $P_1,P_2$ respectively. We note that the probability distribution of $P_0$'s view and $\mathcal{S}_{\rm trunc}^{\rm P_0}$'s output are identical, $P_1$ and $P_2$'s view and $\mathcal{S}_{\rm trunc}^{\rm P_1,P_2}$'s output are identical. Therefore we have the following equations:

\begin{small}
\begin{equation*}
\begin{aligned}
 &\mathcal{S}_{\rm trunc}^{\rm P_0}(\langle z \rangle_0,\langle z \rangle_3,\langle z' \rangle_0,\langle z' \rangle_3) \cong 
    \textbf{view}_{P_0}^{trunc}(\langle z \rangle_k,\langle z' \rangle_k,k \in\{0,1,2,3\}) \\
&\mathcal{S}_{\rm trunc}^{\rm P_1,P_2}(\langle z \rangle_1,\langle z \rangle_2,\langle z' \rangle_1,\langle z' \rangle_2) \cong  \textbf{view}_{P_1,P_2}^{trunc}(\langle z \rangle_k,\langle z' \rangle_k,k \in\{0,1,2,3\})
\end{aligned}
\end{equation*}
\end{small}

\section{Accuracy evaluation over more complex datasets}
\label{sec.accuracy}

We evaluate the accuracy of typical machine learning models, including linear regression, logistic regression, and BP neural networks, trained with \pmpl on more complex datasets, which are Fashion-MNIST and SVHN. (1) Fashion-MNIST is a dataset similar to MNIST.  It also contains 60,000 training samples and 10,000 test samples. Each sample is a 28 $\times$ 28 grayscale image. Rather than handwritten digits as MNIST, Fashion-MNIST contains image samples of ten classes of clothing. (2) SVHN is a dataset from house numbers in Google Street View images. It incorporates more samples, i.e. 73,257 training samples and 26,032 test samples. Besides, each sample is a 32 $\times$ 32 RGB image, associated with a label from ten classes. Furthermore, lots of the images contain some distractors at the sides. Therefore, SVHN and Fashion-MNIST are both harder to classify than MNIST.     
The basic information of these datasets is shown in Table~\ref{table.dataset}.

\begin{table}[h]
\centering
\caption{Brief description of datasets used in \pmpl.}
\label{table.dataset}
\renewcommand\arraystretch{1}
\scalebox{0.9}{
\begin{tabular}{cccc}
\toprule
Dataset       & Fetures & Training samples & Test samples \\ \hline
MNIST         & 784     & 60,000           & 10,000       \\ 
Fashion-MNIST & 784     & 60,000           & 10,000       \\ 
SVHN          & 3,072    & 73,257           & 26,032       \\ \bottomrule
\end{tabular}}
\end{table}

We conduct a series of experiments to compare the accuracy of machine learning models trained with \pmpl and models trained with plaintext decimal data.
As is shown in Table~\ref{table.accuracy}, 
the experimental results show that the accuracy of the machine learning models trained with \pmpl is almost the same as those trained from the data in plaintext. Note that the accuracy of the models of linear regression and logistic regression on SVHN is very poor (about 20\% both in \pmpl and plaintext), thus not shown in Table~\ref{table.accuracy}. In addition, the accuracy of BP neural networks on SVHN is about 73\%, much lower than the result (about 99\%~\cite{DBLP:conf/iclr/ForetKMN21}) from the state-of-the-art neural networks. Thus, we argue that although \pmpl presents a feasible framework with a privileged party, we should pay much attention to enabling \pmpl to efficiently support the state-of-the-art deep neural networks in future.

\begin{table}[ht]
\centering
\caption{Accuracy of the typical machine learning models trained with \pmpl (in secret shares) compared to the ones trained from the decimal data in plaintext.}
\label{table.accuracy}
\renewcommand\arraystretch{1}
\scalebox{0.9}{
\begin{tabular}{cccc}
\toprule
\multirow{2}{*}{Model}                                                         & \multirow{2}{*}{Dataset} & \multicolumn{2}{c}{Accuaracy}           \\ \cline{3-4}
                                                                               &                          & \multicolumn{1}{c}{\pmpl}   & Plaintext \\ \hline
\multirow{2}{*}{\begin{tabular}[c]{@{}c@{}}Linear\\ regression\end{tabular}}   & MNIST                    & \multicolumn{1}{c}{85.77\%} & 85.80\%   \\ 
                                                                               & Fashion-MNIST            & \multicolumn{1}{c}{80.69\%} & 80.80\%    
                                          \\ \hline
\multirow{2}{*}{\begin{tabular}[c]{@{}c@{}}Logistic\\ regression\end{tabular}} & MNIST                    & \multicolumn{1}{c}{91.07\%} & 91.38\%   \\ 
                                                                               & Fashion-MNIST            & \multicolumn{1}{c}{83.99\%} & 84.01\%   
                                         \\ \hline
\multirow{3}{*}{\begin{tabular}[c]{@{}c@{}}BP neural\\ networks\end{tabular}}  & MNIST                    & \multicolumn{1}{c}{96.41\%} & 96.52\%   \\  
                                                                               & Fashion-MNIST            & \multicolumn{1}{c}{86.47\%} & 86.78\%   \\ 
                                                                               & SVHN                     & \multicolumn{1}{c}{73.31\%} & 73.35\%   \\ \bottomrule
\end{tabular}
}
\end{table}

\end{sloppypar}
\end{document}